\documentclass[12pt]{article}
\usepackage{epsfig,graphicx,color}
\usepackage{latexsym}
\usepackage{amstext}
\usepackage{amsmath,amsfonts,amssymb}
\usepackage[normalem]{ulem}
\usepackage{subfigure}
\usepackage{epsfig, graphicx}
\usepackage{color}
\usepackage{amsfonts,bm,color}
\usepackage{verbatim}
\usepackage{float}
\usepackage{latexsym}
\usepackage{bigints}
\usepackage[english]{babel}

\newtheorem{theorem}{Theorem}
\font\bb=msbm10 at 12pt

\def\eE{\hbox{\bb E}}

\begin{document}
%%%%%%%%%%%%%%%%%%%%%%%%%%%%%%%%%%%%%%%%%%%%%%
\title{Stochastic model of a pension plan}
\author{Paz Grimberg and Zeev Schuss\footnote{email: pazgrimberg@gmail.com, schuss@post.tau.ac.il}\\
Department of Applied Mathematics, Tel-Aviv University, \\ Ramat-Aviv, Tel-Aviv 69978,
Israel}
\date{\today}

\maketitle
\begin{abstract}
Structuring a viable pension plan is a problem that arises in the study of financial contracts
pricing  and bears special importance these days. Deterministic pension models often rely on projections that are based on several assumptions concerning the "average" long-time behavior of the stock market. Our aim here is to examine some of the popular "average" assumptions in a more realistic setting of a stochastic model. Thus, we examine the contention that investment in the stock market is similar to gambling in a casino, while purchasing companies, after due diligence, is safer under the premise that acting as a holding company that wholly owns other companies avoids some of the stock market risks. We show that the stock market index faithfully reflects its companies' profits at the time of their publication. We compare the shifted historical dynamics of the S\&P500's aggregated financial earnings to its value, and find a high degree of correlation. We conclude that there is no benefit to a pension fund in wholly owning a super trust. We verify, by examining historical data, that stock earnings follow an exponential (geometric) Brownian motion and estimate its parameters. The robustness of this model is examined by an estimate of a pensioner's accumulated assets over a saving period. We also estimate the survival probability and mean survival time of the accumulated individual fund with pension consumption over the residual life of the pensioner.
\end{abstract}
PACS numbers: 87.10, 89.65.-s,, 89.65G\\
MOS numbers: 91Bxx, 91B60, 91B62, 91B70, 91B28\\
Keywords: Stochastic modeling, market model, pension plan, portfolio, investment
\section{Introduction}
This paper is based on the dissertation \cite{paz}, which contains many additional details of numerical, analytical, and statistical computations of the models discussed below.
\subsection{What is a pension plan?}
A pension plan is a method for a prospective retiree to transfer part of his or her current income stream toward a retirement income.
Pension plans are usually classified into two categories,
\begin{enumerate}
\item A defined-benefit plan - the pension fund (e.g., employer) guarantees the pensioner a fixed,
predefined, benefits upon retirement, regardless of the investment's performance.
\item A defined-contribution plan - the pension fund makes predefined contributions,
usually tax exempt, toward a pool of funds, set aside for the pension fund's future
benefit. The pool of funds is then invested on the retiree's behalf allowing her/him to receive benefits upon retirement. The final  benefit received by the retiree depends on the investment's performance.
\end{enumerate}
The benefits are paid, usually in a lump sum, upon the pensioner's retirement. However,
in some countries, such as the UK, members are legally required to purchase an annuity, which then provides a regular income.

Pensions have a long history in Western civilization. The notion of pension dates back to the Roman Empire \cite{pension_evolution}, where rulers and parliaments provided pensions for their workers, who helped perpetuate their regimes. More than two thousand years ago, the fall of the Roman republic and the rise of the empire were inextricably linked to the payment, or rather the nonpayment, of military pensions. The first private pension was established in 1875 by the American Express Company in the United States \cite{pension_history}. Prior to 1870, private-sector plans did not exist, primarily because most companies were small, family-run enterprises.

\subsection{Public and private pension funds}
A public pension fund is one that is regulated under public-sector law, while a private pension fund is regulated under private-sector law. In certain countries the distinction between public or government pension funds and private pension funds may be difficult to assess. In others, the distinction is made sharply in the law, with very specific requirements for administration and investment. For example, local governmental bodies in the United States are subject to laws passed by the states, in which those localities exist and these laws include provisions, such as defining classes of permitted investments and a minimum municipal obligation \cite{public_vs_private_pension_funds}.
\subsection{The pension crisis }
The obligation of a fixed, predefined amount of benefits upon retirement exposes the insurer to a great risk. The calculation of the the benefits amount is based on financial assumptions that are hard to measure or predict. These assumptions include the lifespan of employees, returns earned by pension investments, future taxes, and rare events, such as natural disasters.

On the other hand, defined-contribution plans transfer the risk to the insured, who is dependent on the pension fund performance upon his/her retirement date. An individual that has retired in 2009 received significantly less than he/she would have in 2011. The following data paint a grim picture of UK and US pension plans.

 In the US, there was a \$1 trillion gap at the end of the fiscal year 2008 between the \$2.35 trillion that American states had to set aside to pay for their employees' retirement benefits and the \$3.35 trillion price tag of those promises \cite{pension_grim_fact1}. The present value of unfunded obligations under Social Security as of August 2010 was approximately \$5.4 trillion \cite{pension_grim_fact5}.
Moreover,American state and local pension plans exhibit a structural shortfall that will likely pose a long-enduring problem, according to the US Congressional Budge Office \cite{pension_grim_fact4}.
In the UK, many employees face retirement with an income well short of their expectations. Employees who pay into a defined-contribution plan for 40 years, may get only half of the retirement income they could have expected \cite{pension_grim_fact2}.
According to the International Monetary Fund \cite{pension_grim_fact3}, Western economies would have to set aside an additional 50\% of their 2010 GDP to support the retirees. Several reforms have been suggested to amend the pension crisis.

\subsection{Reforms}
Reform proposals can be classified into three.
\begin{enumerate}
\item To meet pre-existing defined benefit obligations, the retirement age should be raised.
\item To mitigate risk and reduce obligations, there should be a shift from
defined-benefit to defined-contribution pension plans.
\item To improve accumulated wealth, there should be an increase in resource allocation to fund
pensions by increasing contribution rates and taxes.
\end{enumerate}
The first reform does not exhibit any structural solution, but rather tries to put out a fire. The
second reform still contains the risks of defined-contribution plans, and the third reform involves
raising taxes, which potentially reduces the reward of work and therefore of the incentive to work.

Setting up Super Trusts is based on the belief that its investment policies achieve low-volatility, low-risk, and steady growth. These policies include 100\% stake purchases in companies that produce basic products or commodities, with an underlying economic substance and a high, stable demand. The Super Trusts refrain from investments that carry no fundamental value.
Investments that are considered risky by the Super Trusts are stocks, bonds, currencies, arbitrage-trading, futures, options, and all forms of derivatives.
The motivation for this approach is based on a long list of historical financial crises. To name but a few:
\begin{enumerate}
\item Black Monday (1987) - Dow Jones Industrial Average dropped 22.61\% in one day \cite{black_monday}.
\item Saving and Loans (1980's, 1990's) - Nearly 25\% of all Saving and Loans associations in the United States, worth \$402 billion,  failed \cite{SL_crisis}. The estimated crisis cost in 1996 alone, was \$160 billion in 1996, with a total cost of \$370 billion, 92\% taken from tax payers.
\item Russian Financial Crisis (1998) - Several factors, such as artificially high fixed exchange rate and chronic fiscal deficit led the Russian government to devalue the Ruble, default on domestic debt, and decalre a moratorium on payment to foreign creditors \cite{russian_crisis}. As a result, inflation reached 84\% that year. Banks closed down. Millions of people lost their life savings. As a direct consequence, US Hedge funds collapsed, including Long Term Capital Management (LTCM), which received a \$3.6B bailout \cite{ltcm_bailout}, under the supervision of the Federal Reserve. \
\item The Dot-com bubble (2000) - The NASDAQ Composite lost 78\% of its value. \$5 trillion loss in the market value of companies.
\item The Subprime Mortgage Crisis (2008) - Americans lost more than a quarter of their net worth. Housing prices dropped, GDP began contracting, unemployment rate rose from 5\% to 10\%. S\&P500 fell 57\% from its October 2007 peak. US total national debt rose form 66\% GDP pre-crisis to over 103\% post-crisis \cite{subprime_crisis}.
\end{enumerate}
In addition, the rise of algorithmic trading, systematic trading, high frequency trading, and hedge funds gave birth to new type of stock market crashes - computer code crashes. For example, the 2010 Flash Crash \cite{flash_crash} , and the Knight Capital Group crash \cite{knight_crash} in 2012, are results of crashing of computers running complex algorithms.
These crises led the Super Trusts to seek growth in the net income instead of stock market returns.
\subsubsection{The S\&P500 index as a model of the pensioners' assets}
For the purpose of pension-fund modeling we adopt the S\&P500 index methodology \cite{sp500_methodology} as a representative strategy of a pension fund investment policy. Specifically, the S\&P500 eligibility criteria are
\begin{enumerate}
\item Market value of more than \$4.6B.
\item Annual dollar value traded is greater than its market value in the 6-months period prior to inclusion.
\item At least 250,000 of its shares are traded each month in the 6-months period prior to inclusion.
\item It is a US company.
\item At least 50\% of the company's shares are offered to the public
\item At least 4 consecutive quarters of positive earnings prior to inclusion.
\item Hasn't been initially offered to the public (IPO) for the past 6-12 months.
\end{enumerate}
A company is excluded from the fund if one of the following holds.
\begin{enumerate}
\item It is involved in a merger or acquisition (M\&A) that causes at least 1 violation of the above eligibility criteria.
\item The company is violating at least 1 of the above eligibility criteria on an ongoing basis.
\end{enumerate}

This investment methodology is aligned with purpose of the pension fund, because companies that meet the above requirements are profitable by definition: they exhibit 4 consecutive quarters of positive earnings and moreover, do not exhibit negative earnings on an ongoing basis. Furthermore, these companies are highly liquid and are worth more than \$4.6B. In addition, the historical fact that 97\% of removals from S\&P500 are due to M\&As and that the average time a company stays in the index is 16.7 years, strengthens the notion that these companies are generally profitable and stable. Consequently, these characteristics qualify them as companies with substantial economic substance, suitable for investments by a pension fund.
\subsubsection{The investment performance}
The pension funds are the beneficiaries of the companies' net profit and it is up to the management's discretion to determine the amount of the net profit retained by its underlying companies and the amount accumulated into the pension fund.  In order to be able to gauge fund's performance, we assume that none of the portfolio's net profits is retained by the underlying companies and that the fund accumulates the entire portfolio's net earnings. Therefore, the growth of the fund net income, at time $t$, relative to initial time $t_0$, is given by
$$R(t)=\frac{\sum\limits_{i=1}^{500}NI_i(t)}{\sum\limits_{i=1}^{500}NI_i(t_{i_0})},$$ where $NI_i(t)$ is the net income of the $i$-th company in S\&P500 at time $t$ and $t_{i_0}$ is the time it was first included in the index.

\subsubsection{A refinement of the investment strategy}
\begin{figure}
\begin{center}
\begin{tabular}{c}
	  \includegraphics[width=0.4\textwidth,height=0.25\textheight]{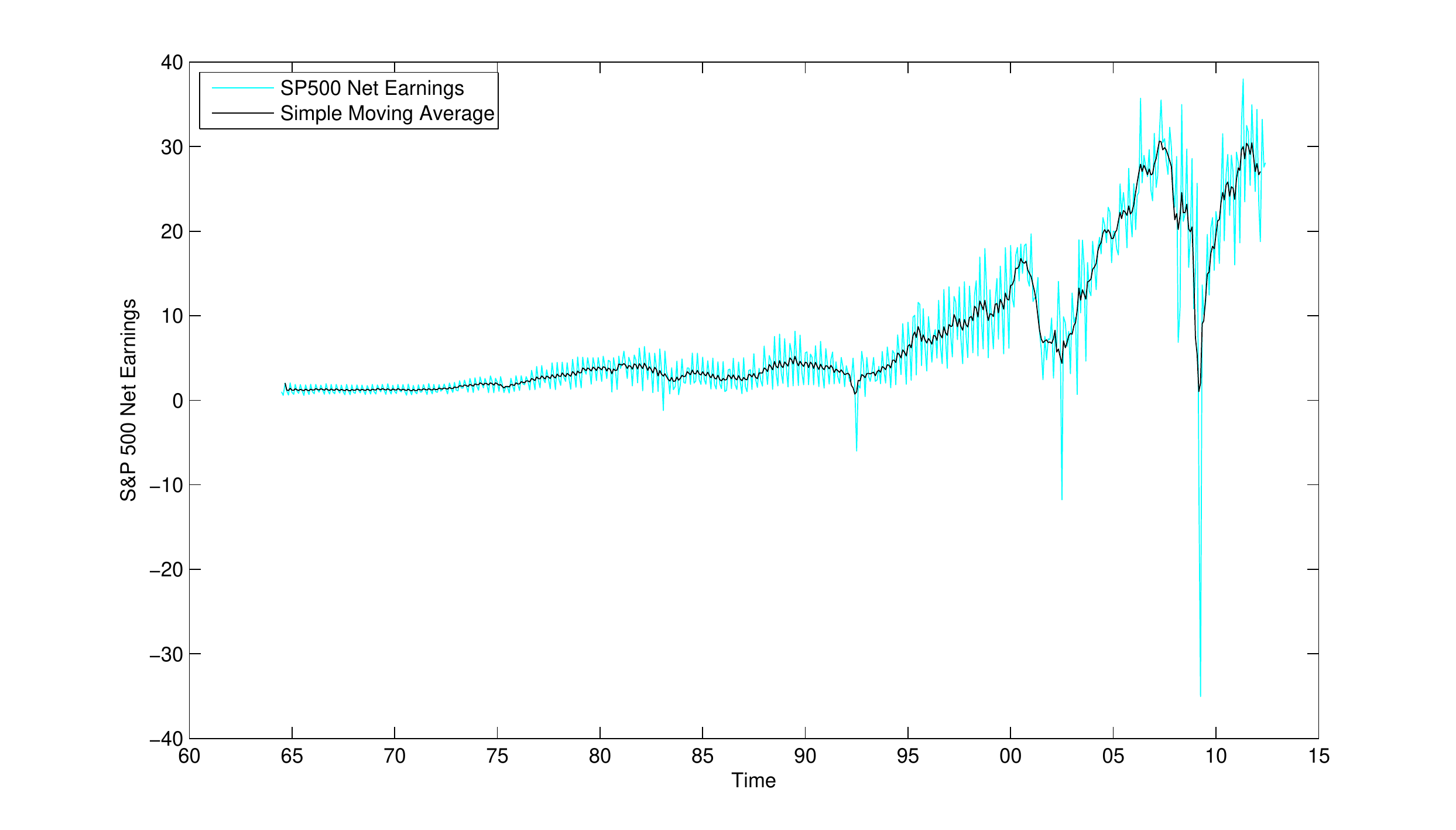}
	  \includegraphics[width=0.4\textwidth,height=0.25\textheight]{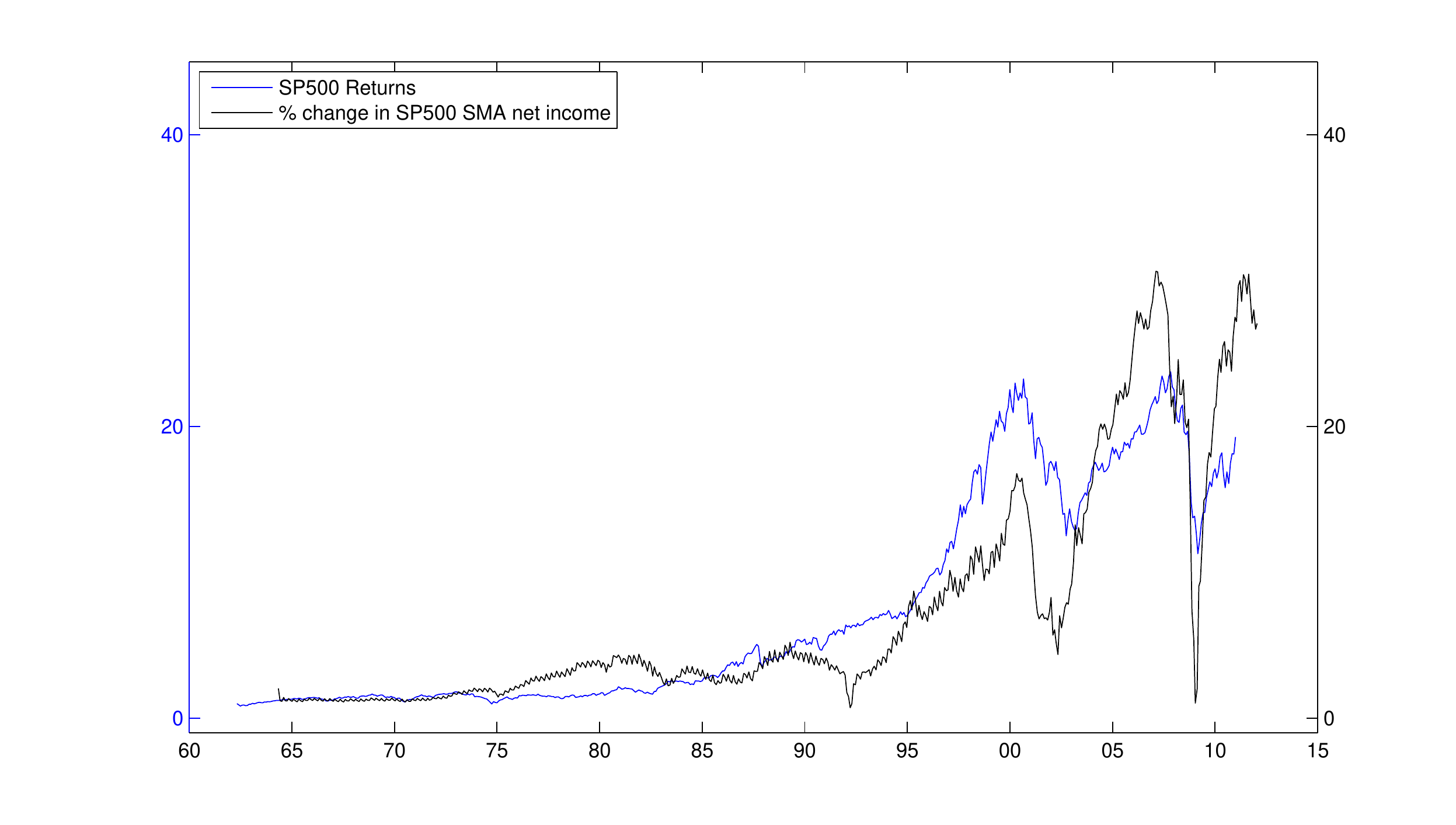}
\end{tabular}
\end{center}
\caption{\small{\bf (Left) } S\&P500 CPI-adjusted net income growth, $R(t)$ (cyan). Moving average of $R(t)$ (black). {\bf (Right)} S\&P500 CPI-adjusted returns (cyan). Moving average of S\&P500 CPI-adjusted net earnings growth (black).}	  \label{fig:sp500}
\end{figure}

\begin{figure}
\begin{center}
\begin{tabular}{c}
	  \includegraphics[width=0.4\textwidth,height=0.25\textheight]{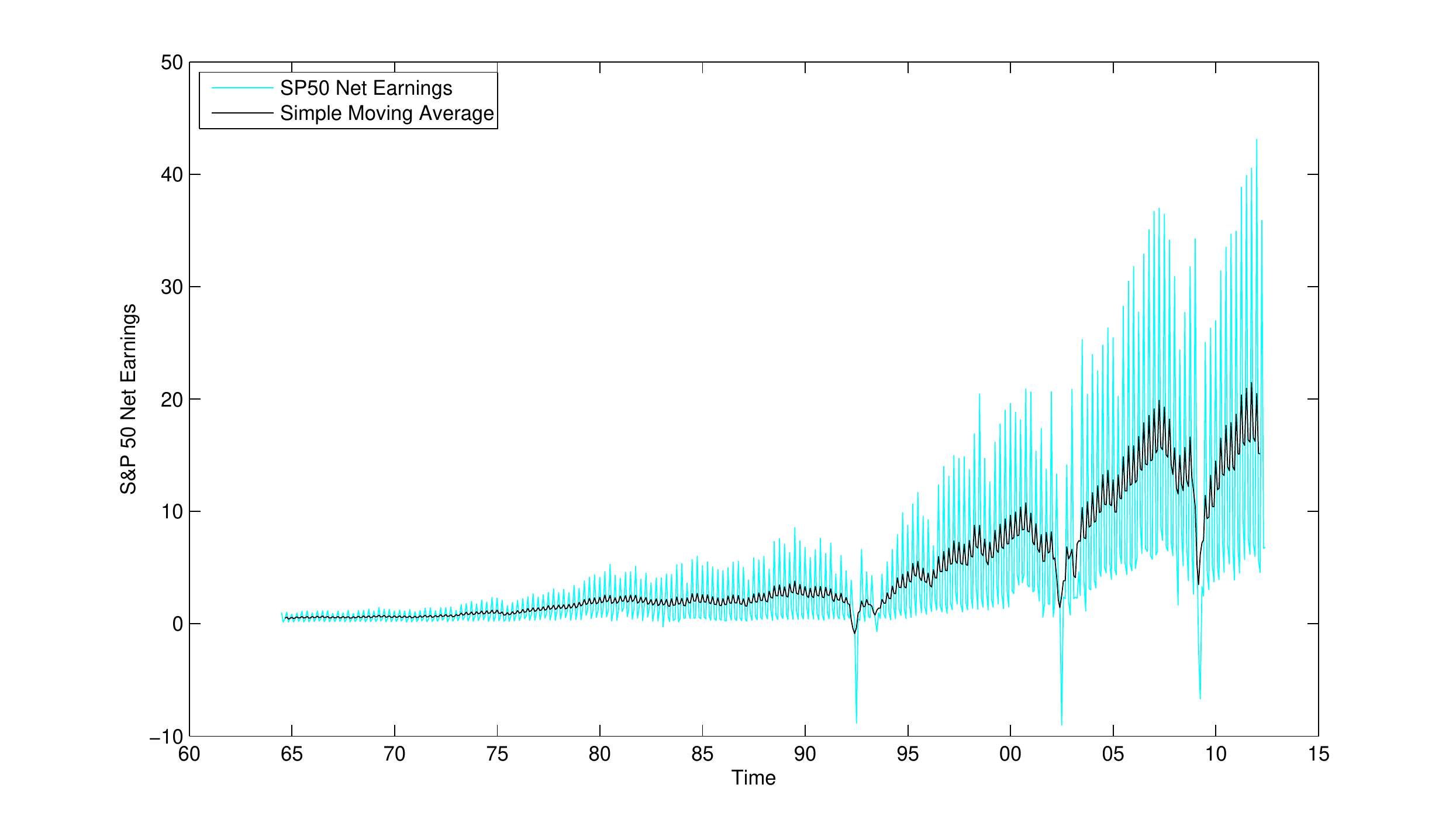}
	  \includegraphics[width=0.4\textwidth,height=0.25\textheight]{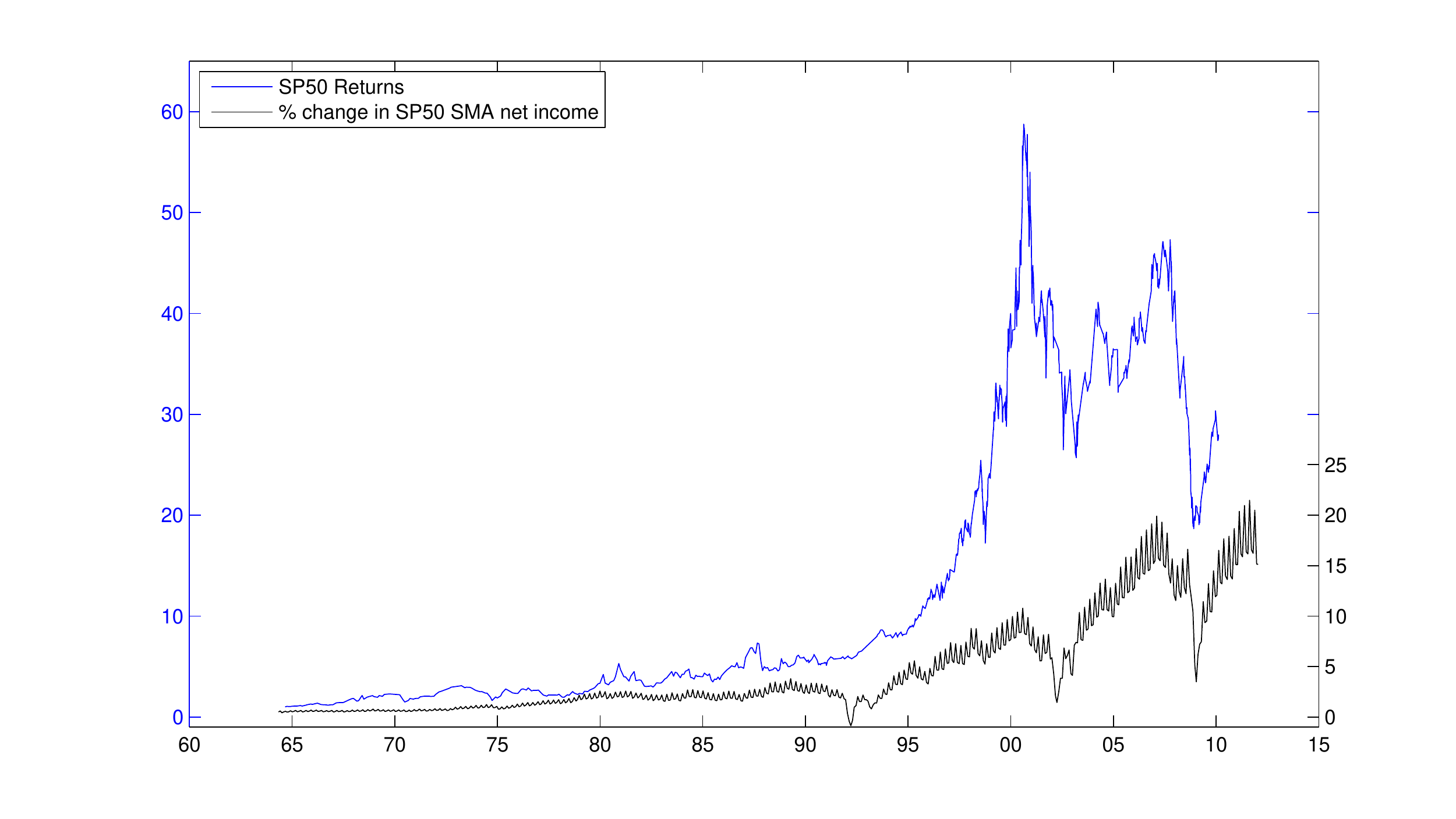}
\end{tabular}
\end{center}
\caption{\small {\bf (Left) Cyan } S\&P50 CPI-adjusted net income growth, $R(t)$. {\bf (Left)
Black } Moving average of $R(t)$. {\bf (Right) 	Cyan } S\&P50 CPI-adjusted returns. {\bf (Right) Black } Moving average of S\&P50 CPI-adjusted net earnings growth.}
	  \label{fig:sp50}
\end{figure}

Fama's Efficient-Market Hypothesis (EMH) was introduced in \cite{fama}. The hypothesis states that it is impossible to "beat the market," because stock market efficiency causes existing share prices to always incorporate and reflect all relevant information. According to the EMH, stocks always trade at their fair value on stock exchanges, making it impossible for investors to either purchase undervalued stocks or sell stocks for inflated prices. Therefore, it is reasonable to assume that the collective price of a market as a whole, as represented by S\&P, for example, should incorporate and reflect all information about its constituents, including their earnings performance at the time of their publication.
We assert this argument by comparing historical returns of two S\&P indices  against their shifted approximated by adding 3 months to the quarter of which the financial statements refer to. For example, if a company reported net income for the 2nd quarter, we assign September 30th as the publication date. The period of 3 months shift was chosen, because US companies are required by law to publish their quarterly financial reports by the end of the subsequent quarter. We obtain financial statements and price data from the CRSP/COMPUSTAT merged database \cite{crsp} and plot the historical monthly performance of their shifted CPI-adjusted net income growth, $R(t)$, from 1970 through 2011. We also plot a 0.6\% window moving average to highlight their trend. To emphasize the high correlation of their dynamics, we plot CPI-adjusted returns performance against the shifted trend of the net earnings curve. The chosen indices representing the market are S\&P50 and S\&P500 (see figures \ref{fig:sp500}, \ref{fig:sp50}).

In addition, we calculate the Pearson correlation coefficient between the shifted earning (raw, not smoothed) and the return performance. The Pearson coefficient is given by
$$ \rho\left(X,Y\right) = \frac{\mbox{Cov}(X,Y)}{\sigma_X \sigma_Y} = \frac{\eE\left[\left(X - \mu_X\right)\left(Y - \mu_Y\right)\right]}{\sigma_X \sigma_Y}.$$ The results seem to belie
the widely held belief that investment in the stock market is similar to gambling in a casino, while purchasing companies is safe.  We conclude that the stock market index faithfully reflects the companies' profits at the time of their publication, thus strengthening the Efficient Market Hypothesis. Moreover, based on analysis of historical data, stock prices perform better, while being just as safe.

In view  of the above, we refine the suggested investment strategy to purchase the shares of S\&P500 companies instead of a 100\% stake in them. Our model of a defined-contribution plan includes an augmented initial influx, invested in the S\&P500 stock market index.

\begin{table}
\begin{center}
\begin{tabular}{|l|c|}
\hline
Index   & $\rho$ coefficient \\
\hline
S\&P50 & 0.89 \\ \hline
S\&P500 & 0.92 \\ \hline
\end{tabular}
\caption{\small Correlations between the shifted CPI-adjusted earnings performance and CPI-adjusted price performance of market-representative US indices.}
\end{center}
\end{table}
\nopagebreak
\section{A stochastic model and its analysis}
In this section, we propose a continuous-time stochastic model of a pension saving portfolio that is invested in the economy, which is represented by the S\&P500 index. We formulate probabilistic questions about the portfolio's robustness and soundness. These questions are reformulated in terms of an initial and boundary value problems for the Fokker-Planck equation for the joint probability density function of the pension portfolio and the salary growth and are answered by solving the equation numerically.

The current value of the portfolio is modeled after the S\&P500 index, which purports to represent the pensioners' asserts. The continuous contribution to the fund from the insured's salary is also modeled as a stochastic process. The two models are combined into a two-dimensional model for the growth of the pension fund, which depends on the initial salary of the pensioner. We are interested in the probabilities of the benefits payable to the individual. The results of the two-dimensional model are summarized in tables in terms of the model's dimensionless parameters and in three-dimensional plots of numerical results.

In addition, we define a stochastic consumption process that describes the individual's rate of resource consumption. We assume a constant dollar amount rate of a pensioner's annual expense and calculate the survival probability of the pension consumption process, that is, the probability that there will still be pension money left after a given number of retirement years. We also calculate the mean first passage time (MFPT) of the consumption process to $0$, which is the expected time for the pension money to run out. Finally, we assume that the pensioner's life span is randomly distributed, according to a certain density function, and calculate the probability that the pension plan survives the pensioner, that is, the probability that the pensioner will die before consuming all his/her pension money.

\subsection{The diffusion model}
A common practice in modeling asset prices is to express factors, such as the Efficient Market
Hypothesis and randomness, in terms of Markov processes, such as diffusion and jump diffusion
processes \cite{diffusion_application}. Accordingly, we model the salaries and the market's fluctuations as diffusion processes and corroborate their drift and diffusion coefficients against historical data. We use historical price data for the stock market index growth model and historical annual wage data for the salaries model. We use continuous-time approximations to the discrete-time processes.

A vector-valued diffusion process $\mbox{\boldmath$x$}(t)$ is a continuous-time Markov process  with almost surely continuous trajectories that satisfies the following conditions \cite{schuss},
\begin{align}
&\lim_{\Delta t \rightarrow 0} \frac{1}{\Delta t}\eE \left\{ \mbox{\boldmath$x$}\left(t + \Delta t\right) - \mbox{\boldmath$x$}(t) \mid \mbox{\boldmath$x$}(t) = \mbox{\boldmath$x$} \right\} = \mbox{\boldmath$a$}(\mbox{\boldmath$x$},t) \nonumber \\
&\lim_{\Delta t \rightarrow 0} \frac{1}{\Delta t}\eE \left\{ \left[\mbox{\boldmath$x$}\left(t + \Delta t\right) - \mbox{\boldmath$x$}(t)\right]\left[\mbox{\boldmath$x$}\left(t + \Delta t\right) - \mbox{\boldmath$x$}(t)\right]^T \mid \mbox{\boldmath$x$}(t) = \mbox{\boldmath$x$} \right\} =\mbox{\boldmath$\sigma$}(\mbox{\boldmath$x$},t)\label{abxt} \\
&\lim_{\Delta t \rightarrow 0} \frac{1}{\Delta t}\eE \left\{ |\mbox{\boldmath$x$}\left(t + \Delta t\right) - \mbox{\boldmath$x$}(t)|^{2+\delta} \mid \mbox{\boldmath$x$}(t) = \mbox{\boldmath$x$} \right\} = 0, \hspace{0.5em} \mbox{for some}\ \delta > 0.\nonumber
\end{align}
We consider diffusion models that are solutions of It\^o stochastic differential equations (SDE) of the form
\begin{align}
 d\mbox{\boldmath$x$}(t) = \mbox{\boldmath$a$}(\mbox{\boldmath$x$}(t),t)\,dt +\mbox{\boldmath$B$}(\mbox{\boldmath$x$}(t),t)\,d\mbox{\boldmath$w$}(t),\label{SDE}
 \end{align}
where $\mbox{\boldmath$w$}(t)$ is a vector standard mathematical Brownian motions (MBM) \cite{schuss} and
$\mbox{\boldmath$B$}(\mbox{\boldmath$x$}(t),t)$ is a matrix such that
$$\mbox{\boldmath$\sigma$}(\mbox{\boldmath$x$},t)=\frac12\mbox{\boldmath$B$}(\mbox{\boldmath$x$}(t),t)\mbox{\boldmath$B$}^T(\mbox{\boldmath$x$}(t),t).$$

In long-time numerical simulations of \eqref{SDE}, we approximate the continuous trajectories by
solutions of the discrete Euler approximation scheme for (\ref{SDE}) with drift coefficient vector
$\mbox{\boldmath$a$}(\mbox{\boldmath$x$},t)$ and diffusion matrix
$\mbox{\boldmath$B$}(\mbox{\boldmath$x$},t)$, which are estimated from historical empirical
trajectories.

%%%%%%%%%%%%%%%%%%%%%%%%%%%%%%%%%%%%%
\subsubsection{Model simplifications} \label{sec:model_simplification}
%%%%%%%%%%%%%%%%%%%%%%%%%%%%%%%%%%%%%
As mentioned above, the drift $\mbox{\boldmath$a$}(\mbox{\boldmath$x$},t)$ and diffusion matrix
$\mbox{\boldmath$B$}(\mbox{\boldmath$x$},t)$ are obtained from sample averaging of historical data in (\ref{abxt}) and fitting interpolated functions. Instead of approximating the It\^o coefficients by interpolation, we could have assumed that these coefficients are also random in the sense that they depend on the particular trajectory of the driving MBM $\mbox{\boldmath$a$}(\mbox{\boldmath$x$},t)$ or are governed by stochastic equations of their own.

\subsubsection{Summary of the exponential Brownian motion}
The scalar exponential Brownian motion $x(t)$ is a modification of the geometric Brownian motion, defined by the linear It\^o equation
\begin{align}
dx(t)= a(t)x(t)\,dt + b(t)x(t)\,dw(t), \quad
x(0) = x_0, \label{lognormal_sde_x}
\end{align}
where $a(t)$ and $b(t)$ are given continuous function and $w(t)$ is MBM. The solution is found in a straightforward manner to be given by
\begin{align}
x(t)= x_0\exp\left\{ \int\limits_{0}^{t}\left[a(s)-\frac12b^2(s)\right]ds + \int\limits_{0}^{t}b(s)\,dw(s) \right\}. \label{solution_of_lognormal_sde}
\end{align}
Its moments $m_k(t) = \eE x^k(t)$ for all $k>0$ are given by
\begin{align}
m_k(t) = x_0^k \exp\left\{ k\int\limits_{0}^{t}a(s)\,ds + \left(\frac{k^2 - k}{2}\right) \int\limits_{0}^{t} b^2(s)\,ds\right\}. \label{moment}
\end{align}
Thus
\begin{align}
\eE x(t) =& x_0 \exp\left\{\int\limits_{0}^{t}a(s)\,ds\right\} \hspace{0.3em} \nonumber \\
\mathrm{Var}\left[x(t)\right]=& x_0^2\exp\left\{ 2\int\limits_{0}^{t}a(s)\,ds\right\}\left(\exp\left\{\int\limits_{0}^{t}b^2(s)\,ds \right\}-1\right).\label{xi_moments}
\end{align}

Alternatively, the moments of $x(t)$ can be calculated by observing that $x(t)$ has the lognormal distribution $x(t) \sim LN\left(\mu,\sigma^2\right),$ where
\begin{align*}
\mu = \log(x_0) + \int\limits_{0}^{t}\left[a(s)-\frac12b^2(s)\right]\,ds,\quad
\sigma^2 =& \int\limits_{0}^{t}b^2(s)\,ds.
\end{align*}
Therefore the moments are
\begin{align}
\eE\left[x(t)\right] &= e^{\mu + \frac12 \sigma^2} \nonumber \\
&=  x_0\exp\left\{\int\limits_{0}^{t}\left[a(s)- \frac12b^2(s)\right]ds + \frac12\int\limits_{0}^{t}b^2(s)\,ds\right\} = x_0\exp\left\{\int\limits_{0}^{t}a(s)\,ds\right\} \nonumber \\ \nonumber \\
\mathrm{Var}[x(t)] &= \left(e^{\sigma^2}-1\right)e^{2\mu + \sigma^2} \nonumber \\
&= x_0^2\exp\left\{2\int\limits_{0}^{t}a(s)\,ds\right\}\left(\exp\left\{\int\limits_{0}^{t}b^2(s)\,ds\right\} - 1\right).
\end{align}

The solution of the inhomogeneous linear SDE
\begin{align}
dX(t) =\hspace{0.3em} \left[a_1(t)X(t) + a_2(t)\right]\,dt + \left[b_1(t)X(t) + b_2(t)\right]\,dW(t),\quad X(t_0) = \hspace{0.3em} X_0 \label{sde_non_homo}
\end{align}
is given by
\begin{align}
X(t) = H(t)\left[1 + \int\limits_{t_{0}}^{t}\left(\frac{a_2(s) - b_1(s)b_2(s)}{H(s)}\right)\,ds + \int\limits_{t_0}^{t} \frac{b_2(s)}{H(s)} dW(s)\right]. \label{nonhomo_linear_sde_solution}
\end{align}
where $H(t)$ is the solution \eqref{solution_of_lognormal_sde} of the homogeneous SDE
\begin{align}
dH(t) = \hspace{0.3em} a_1(t)H(t)\,dt + b_1(t)H(t)dW(t),\quad
H(t_0) = \hspace{0.3em} X_0, \label{sde_homo}
\end{align}
given by
\begin{align}
H(t) = X_0\exp\left\{\int\limits_{t_0}^{t}\left[a_1(s)-\frac12b_1^2(s)\right]ds + \int\limits_{t_0}^{t}b_1(s)dW(s)\right\}.
\end{align}
\subsection{Stochastic model for long-term stock returns}
The S\&P500 index returns process has only one trajectory, so in order to construct its long-term diffusion model, we represent the index as a weighted average of the underlying individual stock returns. Thus, we begin with modeling the dynamics and fluctuations of the consumer-price-index-adjusted (CPI-adjusted) returns of the S\&P500 constituent stocks. Because stock returns $x_i(t)$ are dimensionless, that is,  measured in percents, we posit that although statistically independent, they are statistically identical. That is, all S\&P500 stock returns $x_i(t)$  are the outputs of a single SDE
\begin{align*}
dx(t)= a(x(t),t)\,dt + b(x(t),t) \, dw(t), \quad x(t_0) = 1.
\end{align*}
Equivalently, $x_i(t)$ can be considered outputs of the identical and independent SDEs
\begin{align}
dx_i(t)= a(x_i(t),t)\,dt + b(x_i(t),t) \, dw_i(t), \quad x_i(t_0) = 1\hspace{0.5em}\mbox{for}\ i=1,2,\ldots, \label{SDEo}
\end{align}
where $w_i(t)$ are independent MBMs.
\subsubsection{Discrete approximation scheme for the drift and diffusion coefficients}
We denote by $S(\tau,x)$ the set of all stock returns in the composition of S\&P500 at the end of month $\tau$, whose price had multiplied $x$ times relative to their index inclusion price. If a stock was included in S\&P500 prior to $\tau$ more than once,  the last inclusion date is taken. The trajectory of the stock return process attributable to the $j$-th stock is denoted by $x_j(t)$. The continuous trajectories of (\ref{SDEo}), discounted by the CPI index, are considered to be approximations to the discrete monthly CPI-adjusted return vectors. Thus the drift and diffusion coefficients \eqref{abxt} of the stock returns are approximated with
\begin{align}
a(x,\tau) =& \hspace{0.4em} \frac{1}{|S(\tau,x)|}\sum\limits_{s \in S(\tau,x)} \left[x_s(\tau + 1) - x_s(\tau)\right] \nonumber \\
\label{ret_diff_coef}\\
b^2(x,\tau) =& \hspace{0.4em} \frac{1}{|S(\tau,x)|}\sum\limits_{s \in S(\tau,x)} \left[x_s(\tau + 1) - x_s(\tau)\right]^2.\nonumber \end{align}
\subsubsection{Drift and diffusion surface interpolation} \label{sec:drift_and_diffusion_surface_interpolation}
The source of historical S\&P500 data is the CSRP/COMPUSTAT$^{\circledR}$ merged database for historical monthly stock prices and historical S\&P500 compositions. We use the US Bureau of Labor Statistics for the historical Consumer Price Index (CPI) values.
We compute \eqref{ret_diff_coef} for $\tau$ between January 1970 and December 2011, to obtain the drifts and volatility surfaces
$$G_a = \left\{\left(x,t,a(x,t)\right)\right\},\quad
G_b = \left\{\left(x,t,b^2(x,t)\right)\right\}. $$
These surfaces are interpolated by projecting them onto the $t$-axis, to obtain
$$G_a [t]=\left\{(x,a(x,t)) \right\}, \hspace{0.5em} \mbox{ for }1970 \leq t \leq 2011 ,$$
$$G_b [t]=\left\{(x,b^2(x,t)) \right\}, \hspace{0.5em} \mbox{ for }1970 \leq t \leq 2011.$$
For each $t$ the planar curves $G_a[t]$, and $G_b[t]$ are interpolated with a linear function $\tilde G_a[t]$ and a quadratic polynomial $\tilde G_b[t]$, respectively. The reassembled planar interpolators form the interpolated surfaces
$$\tilde G_a = \left\{ \left(x,t, \tilde G_a[t](x) \right) \in \mathbb{R}^3 \hspace{0.3em} \Big| \hspace{0.3em} 1970 \leq t \leq 2011 \right\},$$
$$\tilde G_b = \left\{\left(x,t, \tilde G_b[t](x)\right) \in \mathbb{R}^3 \hspace{0.3em} \Big| \hspace{0.3em} 1970 \leq t \leq 2011 \right\}.$$
\subsubsection{Numerical results} \label{sec:numerical_results}
Constructing the interpolators $\tilde G_a[t], \tilde G_b[t]$ in the form
\begin{align}
a(x,t) &= \tilde G_a[t](x) = q(t)x + q_2(t)\nonumber \\
b^2(x,t) &= \tilde G_b[t](x) = r(t)x^2 + r_2(t)x + r_3(t),
\end{align}
we determine $q(t),q_2(t),r(t),r_2(t),r_3(t) \in \mathbb{R}$ by minimizing the residuals in the least square sense
$$\sum_{x} \left[\tilde G_a[t] - G_a[t] \right]^2, \quad \sum_{x} \left[\tilde G_b[t] - G_b[t] \right]^2 ,\quad \mbox{ for every $1970\leq t \leq 2011$.}$$

The interpolating functions, $\tilde G_a[t]$, $\tilde G_b[t]$, plotted against the projections $G_a[t]$, $G_b[t]$, are given in \cite{paz}.
\subsubsection{The fit parameters}
\begin{figure}[ht!]
\begin{center}
	  \includegraphics[width=0.32\textwidth,height=0.32\textheight]{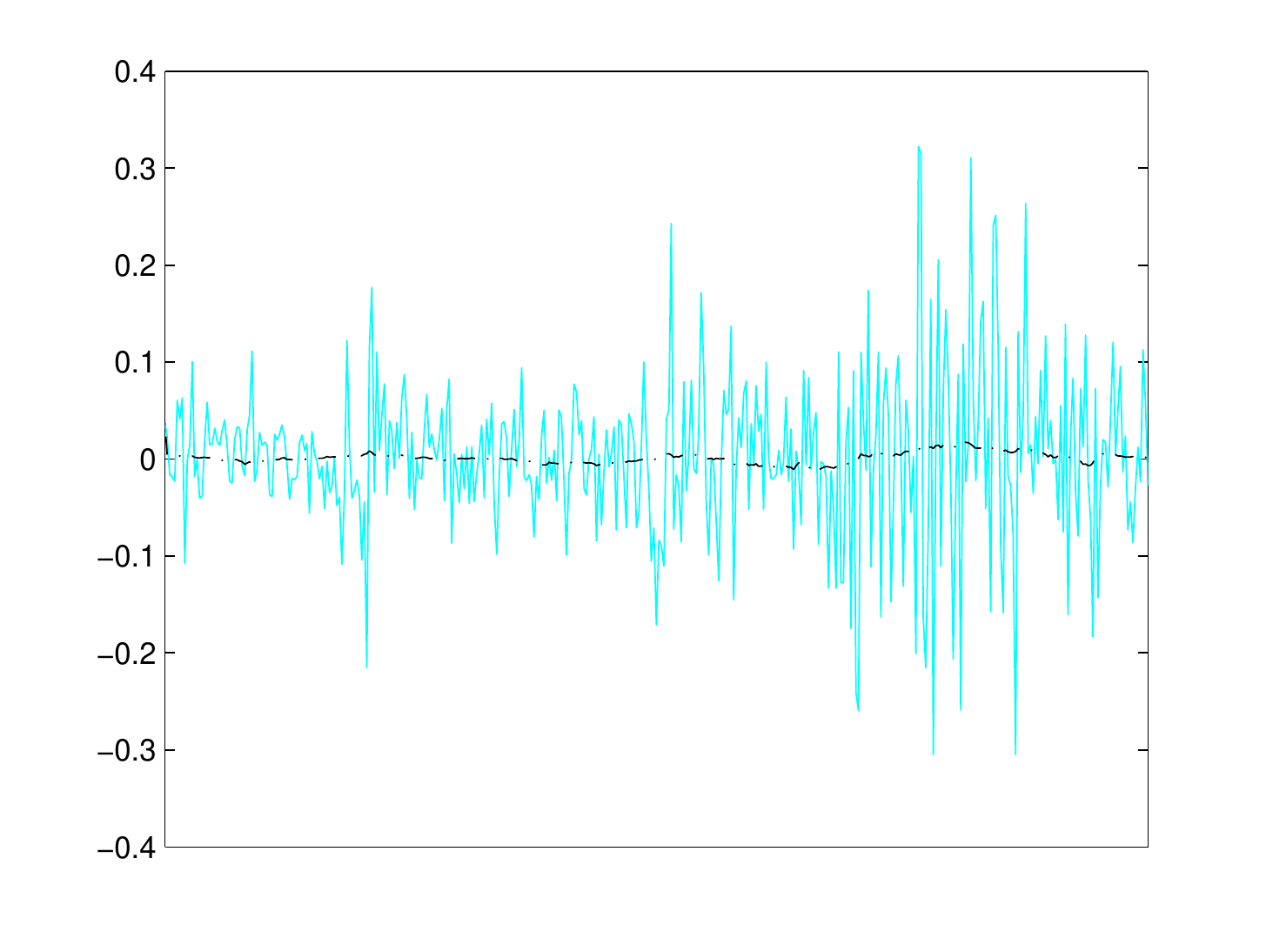}
	  \includegraphics[width=0.32\textwidth,height=0.32\textheight]{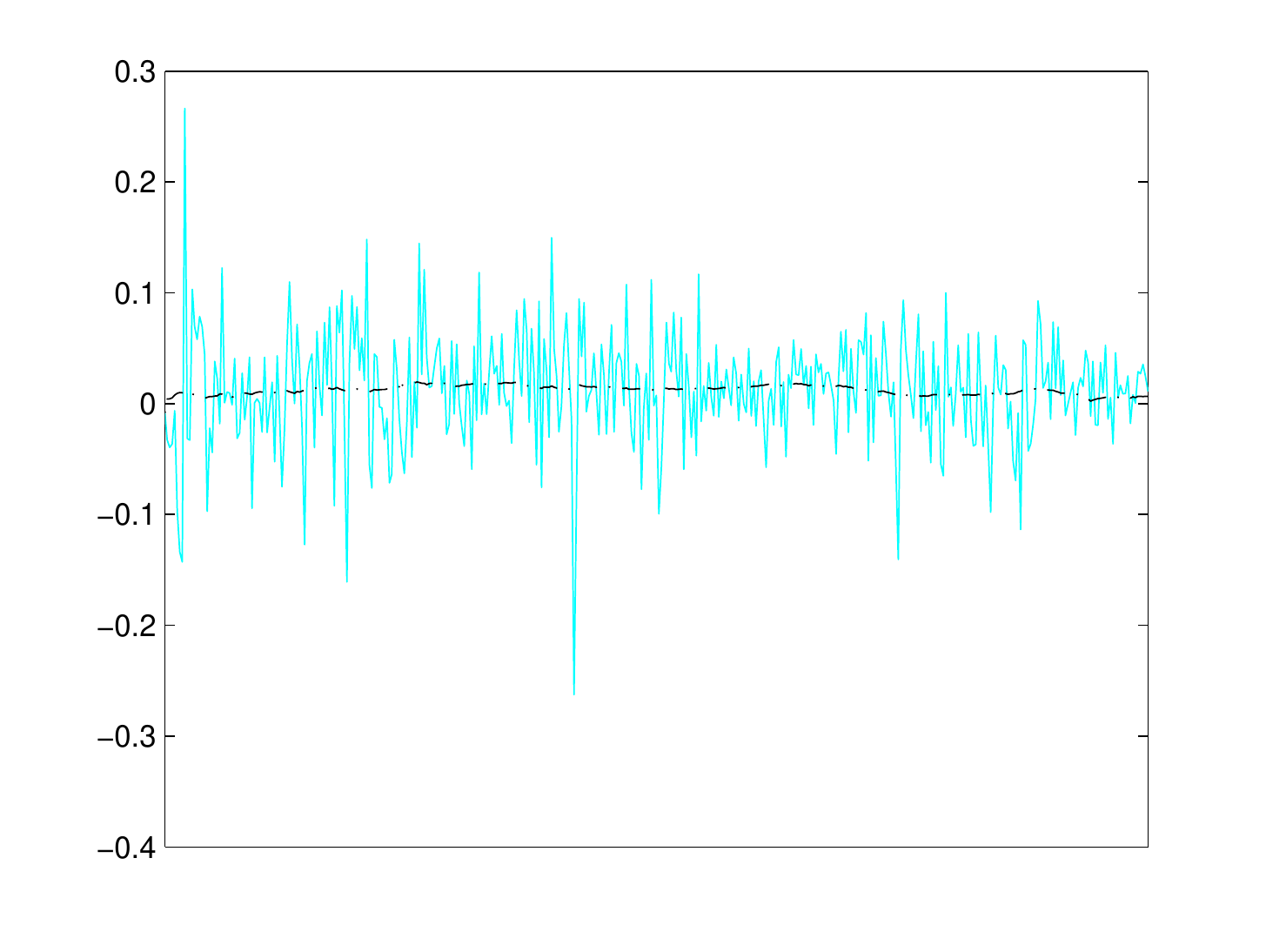}
	  \caption{\small {\bf (Left)} The slopes $q(t)$ between 1970 and 2011 (cyan).  The moving average of the slopes $q(t)$ with a 5\% window (black). {\bf (Right)}  The constant terms $q_2(t)$ between 1970 and 2011 (cyan). The moving average of the constant terms $q_2(t)$  with a 5\% window (black).}
	  \label{fig:a1a2}
\end{center}
\end{figure}

\begin{figure}[ht!]
\begin{center}
	  \includegraphics[width=0.32\textwidth,height=0.32\textheight]{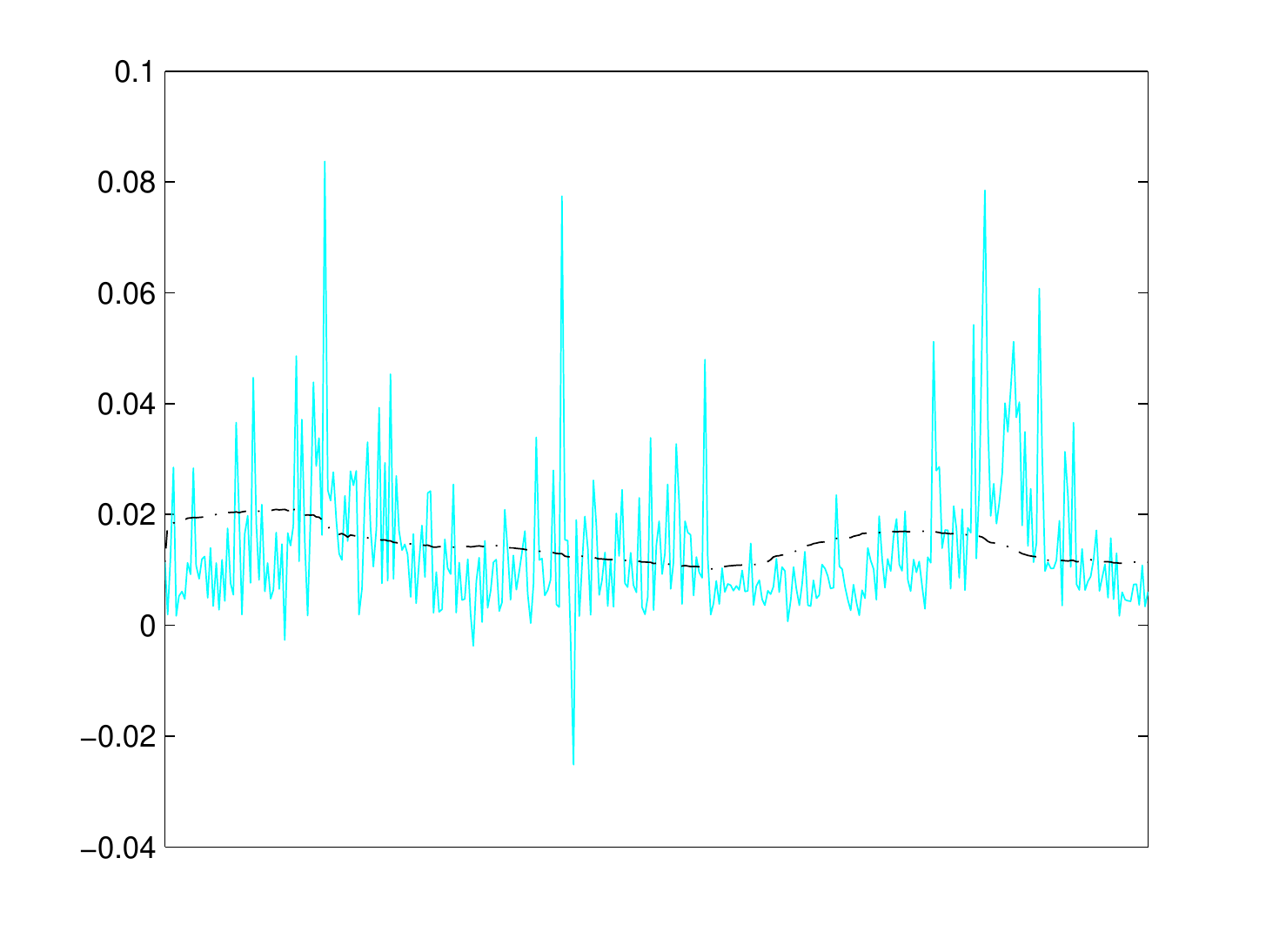}
	  \includegraphics[width=0.32\textwidth,height=0.32\textheight]{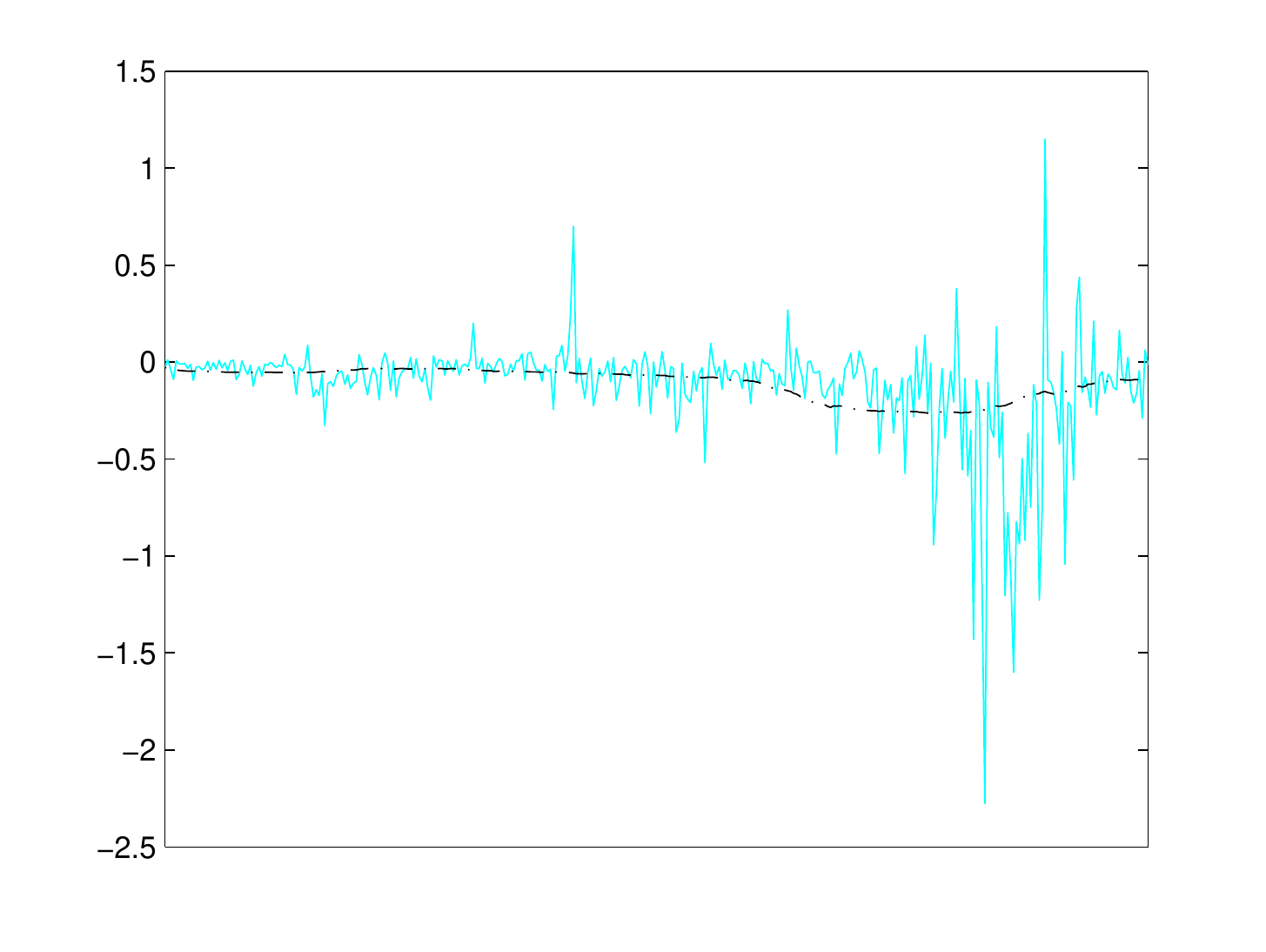}
	  \includegraphics[width=0.32\textwidth,height=0.32\textheight]{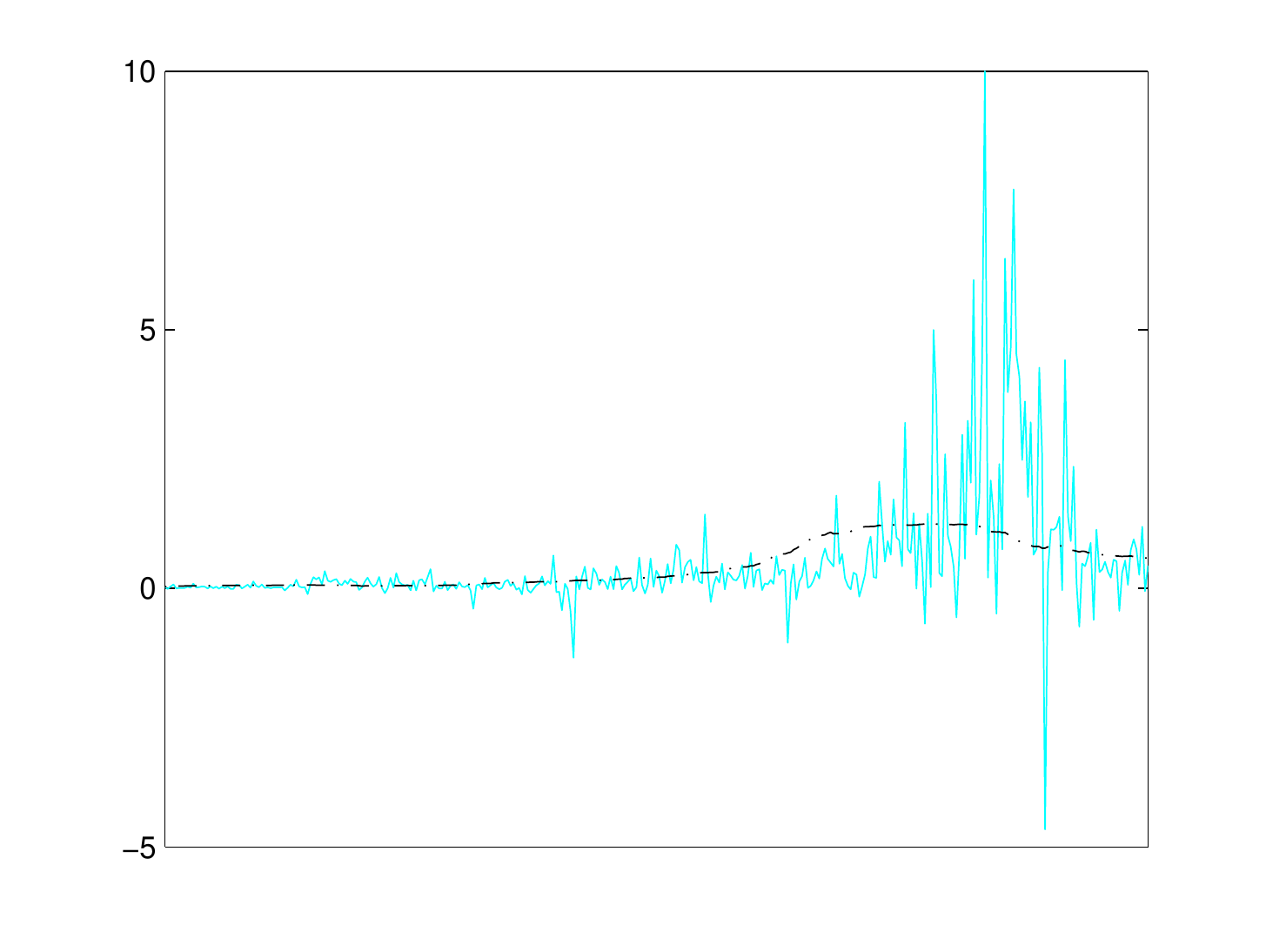}
	  \caption{\small {\bf (Left)} The leading quadratic coefficient $r(t)$ between 1970 and 2011 (cyan).  The moving average of $r(t)$ with a 5\% window (black). {\bf (Middle)}  The  coefficient  $r_2(t)$ between 1970 and 2011 (cyan). The moving average of the constant terms $r_2(t)$ with a 5\% window (black). {\bf (Right)}  The  coefficient  $r_3(t)$ between 1970 and 2011 (cyan). The moving average of $r_3(t)$ with a 5\% window (black). }
	  \label{fig:b1b2b3}
\end{center}
\end{figure}

In figure \ref{fig:a1a2}, the coefficient $q(t)$ is plotted for $1970 \leq t \leq 2011$. For simplicity (see
\ref{sec:model_simplification}), we approximate the function $q(t)$ with its moving average, that is, at every point the function equals the average of its values at the $N$ preceding points.  The resulting approximation is the constant $$q(t)=0.002742.$$
The function $q_2(t)$ is plotted for $1970 \leq t \leq 2011$ and its moving average results in the constant value
$$q_2(t) \equiv 0.$$
In figure \ref{fig:b1b2b3}, the coefficient $r(t)$ is plotted for $1970 \leq t \leq 2011$ and its moving average is the constant value
$$r(t) = 0.01.$$
The function $r_2(t)$ is plotted for $1970 \leq t \leq 2011$ and its moving average is the constant value
$$r_2(t) = 0.$$
The function $r_3(t)$ is plotted for $1970 \leq t \leq 2011$ and its moving average is the constant value
$$r_3(t) \equiv 0.$$
Substituting $q,q_2,r,r_2,r_3$ into \eqref{SDEo}, we get
\begin{align}
dx_i(t) = q x_i(t)\,dt + r x_i(t) dw_i(t), \quad x_i(t_{i_0}) = x_0. \label{x_sde}
\end{align}

\subsubsection{Change of time scale}
The time units of the stochastic process $x_i(t)$ are months. We change to time scale of years, so we can match the time scale of the annual income data.
\newpage

For any positive constant $c$  the process transformation
\begin{align}
w_2(t) = cw(t/c^2) \label{brownian_time_scaling}
\end{align}
is also a MBM \cite{schuss}. Using \eqref{solution_of_lognormal_sde}, the solution of \eqref{x_sde} is given by
\begin{align}
x_i(t) = x_o\exp\left\{ \left(q - \frac12 r^2\right)t + rw_i(t)\right\},
\end{align}
so together with \eqref{brownian_time_scaling} and the value $c=1/\sqrt{12}$, we get $x_i(t)$ for $t$ measured in years as
\begin{align}
x_i(t) = x_0\exp\left\{12\left(q - \frac12r^2\right)t + \sqrt{12}\hspace{0.2em}r w_i(t)\right\}
\end{align}
and it satisfies the stochastic equation
\begin{align}
dx_i(t) =& \hspace{0.4em} \psi x_i(t)\,dt + \phi x_i(t)\,dw_i(t) \nonumber \\
x_i(t_{i_0}) =& \hspace{0.4em} x_0, \label{x_sde_scaled}
\end{align}
with the constants
\begin{align*} \psi = 12q = 0.0329 \mbox{ , } \phi = \sqrt{12}r = 0.3464.
\end{align*}
\subsection{Stochastic model of the stock market returns index}
We next seek to identify the stochastic dynamics of the stock market returns index. The returns index  is determined by the weighted average of its constituents returns. We show below that for a large number of summands, the behavior of a weighted average, under certain assumptions, coincides with the arithmetical mean.
\subsubsection{A weak law of large numbers for weighted averages}
We consider a sequence of i.i.d. random variables $x_i$ with finite first moment $\mu$ and variance $\sigma^2$. For an increasing double sequence of weights $\lambda_{i,n}$ , $i=1,2,\ldots,n$ and $n=1,2 \ldots $ such that $\sum_{i=1}^{n} \lambda_{i,n} = 1$ and $\sum_{i=1}^{n}\lambda_{i,n}^2 = O(n^{-1})$.
The first two moments of the weighted average $X_n(t) = \sum_{i=1}^{n}\lambda_{i,n}x_i$ are given by
\begin{align*}
\eE \left[X_n\right] &= \sum_{i=1}^{n}\lambda_{i,n}\eE x_i = \mu, \nonumber \\
\mathrm{Var}\left[X_n\right] &= \eE \left[ \left(\sum_{i=1}^{n}\lambda_{i,n}x_i\right)^2\right] - \mu^2 = \nonumber \\
&= \sum_{i \neq j } \lambda_{i,n}\lambda_{j,n} \eE[x_i]\eE[x_j] + \sum_{i=1}^{n}\lambda_{i,n}^2\eE[x_i^2] - \mu^2 \nonumber \\
&= \mu^2\sum_{i \neq j } \lambda_{i,n}\lambda_{j,n} + \left(\sigma^2 + \mu^2 \right)\left(\sum_{i=1}^{n}\lambda_{i,n}^2\right) - \mu^2 \nonumber \\
&= \mu^2 \left(\sum_{i=1}^{n}\lambda_{i,n}\right)^2 + \sigma^2\left(\sum_{i=1}^{n}\lambda_{i,n}^2\right) - \mu^2 \nonumber \\
&= \sigma^2 \sum_{i=1}^{n}\lambda_{i,n}^2 = \sigma^2 O(n^{-1}).
\end{align*}
Tchebychev's inequality gives
\begin{align*}
\Pr\left\{\left\vert X_n - \mu \right\vert > \epsilon \right\} \leq \frac{\mathrm{Var}\left[X_n\right]}{\epsilon^2} = \frac{\sigma^2O(n^{-1})}{\epsilon^2} = \sigma^2O(n^{-1}),
\end{align*}
hence, for every $\epsilon>0$
\begin{align*}
\lim_{n \to \infty} \Pr\left\{\left\vert X_n - \mu \right\vert > \epsilon \right\} = \lim_{n \to \infty }\sigma^2 O(n^{-1}) = 0.
\end{align*}
It follows that
\begin{align}
\lim_{n \to \infty} \Pr\left\{ \left\vert \sum_{i=1}^{n}\lambda_{i,n}x_i - \frac1n\sum_{i=1}^{n}x_i \right\vert > \epsilon \right\} = 0. \label{weighted_average_cor}
\end{align}

\subsubsection{Index model with equal weights}
The S\&P500 end-of-year weights from 2001 to 2011 are well approximated with $\lambda_{i,n} = i^\alpha/\sum_{i=1}^{n}i^\alpha$, for $\alpha = 18$ (see \cite{paz}).

Such weights satisfy the weak law of large numbers for a weighted average. Indeed,
\begin{align*}
\lambda_{i,n} = \frac{i^\alpha}{\sum_{i=1}^{n}i^\alpha} = \frac{\frac1n \left(\frac in\right)^{\alpha}}{\sum_{i=1}^{n}\left(\frac{i}{n}\right)^{\alpha} \frac1n } \approx \frac{\frac1n \left(\frac in\right)^{\alpha}}{\int_{0}^{1}x^{\alpha}dx} = \frac{(\alpha+1)}{n}\left(\frac in \right)^{\alpha}
\end{align*}
and
\begin{align*}
\sum_{i=1}^{n} \lambda_{i,n}^2 \approx \frac{(\alpha+1)^2}{n}\sum_{i=1}^{n}\left(\frac in\right)^{2\alpha} \frac1n \approx \frac{(\alpha + 1)^2}{n}\int_{0}^{1}x^{2\alpha}dx = n^{-1} \frac{(\alpha+1)^2}{2\alpha+1} = O(n^{-1}).
\end{align*}
Therefore, by  \eqref{weighted_average_cor}, we assume henceforth an equal-weights index
\begin{align}
X_n(t) = \frac1n\sum\limits_{i=1}^{n}x_i(t).
\end{align}

The drift of the sum of lognormal stochastic processes is linear and therefore equal to the average of the underlying drifts. However, the diffusion coefficient is obtained by assembling $n$ independent MBMs motions into one. Therefore, the SDE of $X_n(t)$, is given by
\begin{align}
dX_n(t) =& d\left(\frac1n\sum\limits_{i=1}^{n}x_i(t)\right) = \psi(t)\frac1n\sum\limits_{i=1}^{n}x_i(t)\,dt + \phi(t)\frac1n\sum\limits_{i=1}^{n}x_i(t)\,dw_i(t) \nonumber \\
=& \psi(t)X_n(t)\,dt + \frac1n\phi(t)\left(\sqrt{\sum\limits_{i=1}^{n}x_i^2(t)}\right)\,dW(t). \label{Xn_SDE}
\end{align}

Much research has been done on the subject of identifying the distribution of the average of lognormal random variables (rvs). Large Deviation Theory and Central Limit Theorem methods tend to fail, because a moment generating function does not exist for lognormal rvs. Several numerical methods have been suggested for the approximation of the sum of lognormal rvs. In \cite{saddlepoint_approx}, the steepest descent technique is used to numerically evaluate the cumulative distribution function (cdf) for a sum of lognormal rvs,  using the Lambert-W function. This method works well for only a few summands with relatively low variance. In our case, where long-term investment is considered, the variance becomes large. In the Fenton-Wilkinson (F-W) method \cite{FW1}, \cite{FW2}  the sum is approximated with another lognormal, whose first two moments are matched to the sum. Numerical simulations show that the F-W method is a good approximation of the average process, for long time periods.
\subsubsection{Lognormal approximation of the pdf of lognormal i.i.d. random variables}
We employ the F-W moment matching technique to construct a linear stochastic equation
\begin{align}
dZ_n(t) = \psi(t)Z_n(t)\,dt + \Phi(t)Z_n(t)\,dw(t). \label{Z_SDE}
\end{align}
such that
\begin{enumerate}
\item $Z_n(0)=X_n(0)$
\item $\eE\left[Z_n(t)\right] = \eE\left[X_n(t)\right]$, for every $t \geq 0$
\item $\mathrm{Var}\left[Z_n(t)\right] = \mathrm{Var}\left[X_n(t)\right]$ , for every $t \geq 0$.
\end{enumerate}
The functions $\psi(t)$ and $\Phi(t)$ are chosen so that condition (3) is satisfied. Employing the formula \eqref{moment} for lognormal moments, we obtain
\begin{align}
\mathrm{Var}\left[Z_n(t)\right] = Z_n^2(0)\exp\left\{2\int_{0}^{t}\psi(s)\,ds\right\}\left(\exp\left\{\int_{0}^{t}\Phi^2(s)\,ds\right\}-1\right). \label{VarZn}
\end{align}
Calculating the variance of $X_n(t)$ directly, we get
\begin{align}
\mathrm{Var}\left[X_n(t)\right] =& \frac{1}{n^2}\mathrm{Var}\left[\sum_{i=1}^{n}x_i(t)\right] = \frac{1}{n^2}\sum_{i=1}^{n}\mathrm{Var}\left[x_i(t)\right] \nonumber \\
=& \frac1n x_0^2\exp\left\{2\int_{0}^{t}\psi(s)\,ds\right\}\left(\exp\left\{\int_{0}^{t}\phi^2(s)\,ds\right\}-1\right). \label{VarXn}
\end{align}
Equating \eqref{VarZn} and \eqref{VarXn}, together with $Z_n(0)=X_n(0)=x_i(0)$, we get
\begin{align}
\exp\left\{\int_{0}^{t}\Phi^2(s)\,ds\right\} = \frac1n\left(\exp\left\{\int_{0}^{t}\phi^2(s)\,ds\right\} + n - 1\right),
\end{align}
hence
\begin{align}
\int_{0}^{t}\Phi^2(s)\,ds = \log\left(\exp\left\{\int_{0}^{t}\phi^2(s)\,ds\right\} + n - 1 \right) -\log n.
\end{align}
Differentiating, we find that
\begin{align}
\Phi^2(t) = \frac{\phi^2(t)\exp\left\{\int_{0}^{t}\phi^2(s)\,ds\right\}}{\exp\left\{\int_{0}^{t}\phi^2(s)\,ds\right\} + n - 1}.
\end{align}
Thus the SDE \eqref{Z_SDE} is given by
\begin{align}
dZ_n(t) = \psi(t)Z_n(t)\,dt + \left(\frac{\phi^2(t)\exp\left\{\int_{0}^{t}\phi^2(s)\,ds\right\}}
{\exp\left\{\int_{0}^{t}\phi^2(s)\,ds\right\} + n - 1}\right)Z_n(t)dW(t),\label{Z_SDE_explicit}
\end{align}
and its solution satisfies conditions (1)-(3). We note that $\Phi^2(t)\longrightarrow \phi^2(t)$ as
$t \rightarrow \infty$, meaning that the asymptotic behaviour of $Z_n(t)$ aligns with that of the underlying stocks of $X_n(t)$. The solution of \eqref{Z_SDE} is given by
\begin{align}
Z_n(t) = x_0\exp\left\{ \int_{0}^{t}\left[\psi(s)-\frac12\Phi^2(s)\right]ds + \int_{0}^{t}\Phi(s)dW(s) \right\} \label{Zn},
\end{align}
and in terms of the underlying stocks,
\begin{align*}
Z_n(t) =& x_0\exp\left\{\int_{0}^{t}\left[\psi(s) - \frac{ \phi^2(s)\exp\left\{\int_{0}^{s}\phi^2(p)\,dp\right\}} {2\left(\exp\left\{\int_{0}^{s}\phi^2(p)\,dp\right\} + n-1\right)}\right]\,ds\right.\\
&\left.+ \int_{0}^{t}\left(\frac{ \phi^2(s)\exp\left\{\int_{0}^{s}\phi^2(p)\,dp\right\} }{\exp\left\{\int_{0}^{s}\phi^2(p)\,dp\right\} + n-1}\right)^{\frac12}\,dW(s) \right\}.
\end{align*}
Finally, we incorporate the estimated coefficients, $\psi(s) = \psi, \phi(t) = \phi$, to get
\begin{align}
Z_n(t) =& x_0\exp\left\{\int_{0}^{t}\left(\psi -\frac{ \phi^2e^{\phi^2s } }{2\left(e^{\phi^2s} + n-1\right)}\right)\,ds + \int_{0}^{t}\left(\frac{ \phi^2e^{\phi^2s } }{e^{\phi^2s} + n-1}\right)^{\frac12}dW(s) \right\} \nonumber \\
=& x_0\exp\left\{\psi t -\frac12\log\left( e^{\phi^2t} + n - 1\right) +\frac12\log(n)  + \int_{0}^{t}\left(\frac{ \phi^2e^{\phi^2s } }{\phi^2e^{\phi^2s} + n-1}\right)^{\frac12}dW(s) \right\} \nonumber \\
=& x_0\left(\frac{e^{\phi^2t} + n - 1}{n}\right)^{-\frac12 }\exp\left\{\psi t + \int_{0}^{t}\left(\frac{ \phi^2e^{\phi^2s } }{\phi^2e^{\phi^2s} + n-1}\right)^{\frac12}dW(s)\right\}. \label{Zn}
\end{align}
\subsubsection{Euler scheme simulations of $X_n(t)$}
The Wiener interpretation of stochastic differential equations is useful for both the conceptual understanding of SDEs and for deriving differential equations that govern the evolution of the pdf's of their solutions \cite{schuss}. It\^o's definition of the stochastic integral on the lattice $t_k = t_0 +  k\Delta t$, with $\Delta t = T/N$ and $\Delta w(t) = \Delta w(t + \Delta t) - w(t)$, defines the solution of the SDE \eqref{lognormal_sde_x}, or equivalently, of the It\^o integral equation
\begin{align}
x(t) = x_0 + \int\limits_{0}^{t}a(x(s),s)\,ds + \int\limits_{0}^{t}b(x(s),s)\,dw(s), \label{ito_integral}
\end{align}
as the limit of the solution of the Euler scheme
\begin{align}
x_N(t+\Delta t) = x_N(t) + a(x_N(t),t)\Delta t + b(x_N(t),t)\Delta w(t),\quad
x_N(0) = x_0 \label{euler_scheme}
\end{align}
as $\Delta t \to 0$. The increments $\Delta w(t)$ are independent random variables that can be constructed by Levy's method \cite{schuss}, as $\Delta w(t) = n(t)\sqrt{\Delta t}$, where the random variables $n(t)$, for each $t$ on the numerical mesh, are independent standard Gaussian rvs $\mathcal{N}(0,1)$. according to the recursive scheme \eqref{euler_scheme}. At any time $t$ on the numerical mesh, the process $x_N(t)$ depends on the sampled trajectory $w(s)$ for $s \leq t$, so it is $\mathcal{F}_t$-adapted. The existence of the limit $x(t) = \lim\limits_{N \to \infty} x_N(t)$  is guaranteed by the following theorem
\begin{theorem}[{\bf Skorokohd \cite{skorokhod}}]
If $a(x,t)$ and $b(x,t)$ are uniformly Lipschitz  continuous functions in $x\in\mathbb{R}$, $t\in[t_0,T]$, then the limit $x(t) \stackrel{\Pr}=\lim\limits_{N \to \infty} x_N(t)$ (convergence in probability) exists and is the solution of \eqref{ito_integral}.
\end{theorem}
The convergence of the pdf is guaranteed by the theorem
\begin{theorem}[{\bf \cite{schuss}}]
The pdf $p_N(x,t \mid x_0)$ of the solution $x_N(t,\omega)$ of \eqref{euler_scheme} converges  as $N \to \infty$ to the solution $p(x,t \mid x_0)$ of the FPE $$\frac{\partial p(y,t \mid x,s}{\partial t} = \frac12 \frac{\partial^2\left[b^2(y,t)p(y,t \mid x,s)\right]}{\partial y^2} - \frac{\partial\left[a(y,t)p(y,t \mid x,s)\right]}{\partial y}$$ with the initial condition $\lim\limits_{t\hspace{0.1em}\downarrow \hspace{0.1em} s}p(y,t\mid x,s) = \delta(y-x)$.
\end{theorem}
We construct 10,000 trajectories of $X_n(t)$, by averaging its 500 underlying trajectories $x_i(t)$, for $0 \leq t \leq 600$ months. The trajectories $x_i(t)$ are constructed with the scheme
\begin{align*}
x_i(t) = x_i(t-1) + \phi \hspace{0.2em} x_i(t-1) \cdot \mathcal{N}(0,1),\hspace{0.5em}\mbox{for}\ 1 \leq t \leq 600,\quad x_i(0) = 1.
\end{align*}
The quality of the lognormal fit is shown in \cite{paz}.

\subsection{Stochastic model for salaries}
We assume that the salary growth $s_i(t)$ of member $i$ by time $t$ is governed by the SDE
\begin{align}
ds_i(t)= a(s_i,t)\,dt + b(s_i,t)\,dw_i(t),\quad s_i(t_0) = 1, \label{ds_i}
\end{align}
where $w_i(t)$ are independent MBMs.
\subsubsection{Discrete approximation scheme for the drift and diffusion coefficients}
We approximate the discrete trajectories of the  yearly CPI-adjusted wage vectors by the continuous trajectories of \eqref{ds_i}. We
denote by $S(\tau,x)$ the set of all individuals whose salary at time $\tau$ had multiplied $x$ times relative to their initial
values. The trajectory of the salary growth process, attributable to the $j$-th individual, is denoted by $x_j(t)$. The coefficients of the approximating SDE \eqref{ds_i} are defined by the empirical averages \eqref{ret_diff_coef}.

\begin{figure}[ht!]
\begin{center}
	  \includegraphics[width=0.45\textwidth,height=0.3\textheight]{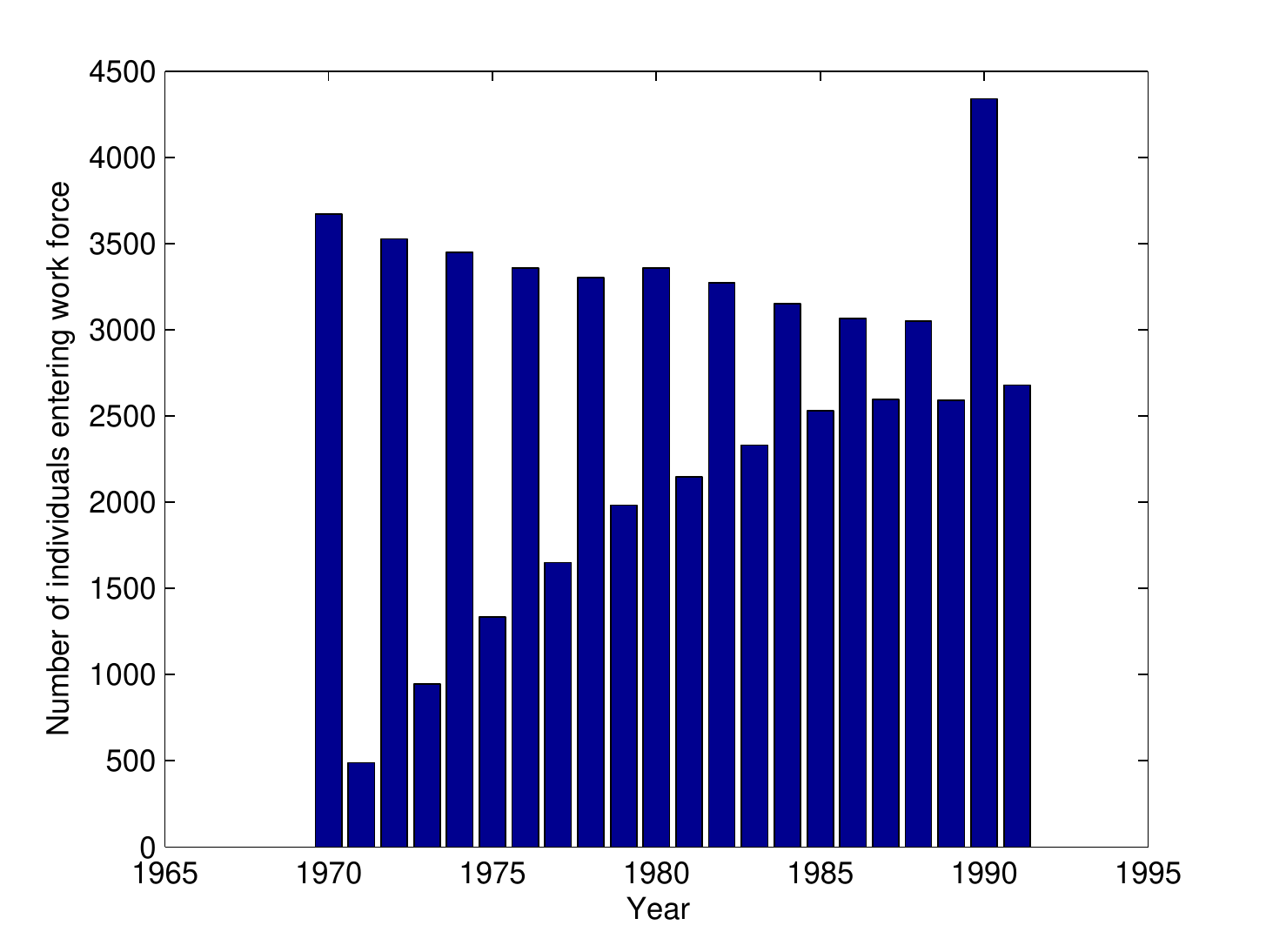}
	  \caption{{\small The number of PSID-surveyed individuals entering work force, by year.}}
	  \label{fig:psid}
\end{center}
\end{figure}

 The data for the construction \eqref{ret_diff_coef} for the wage model are taken from the Panel Study of Income Dynamics$^{\circledR}$ database \cite{psid}. PSID is a longitudinal survey of a representative sample of US individuals and families, which has been taken since 1968. Information on individuals and their descendants has been collected continuously, including data covering employment, income, wealth, expenditures, health, marriage, childbearing, child development, philanthropy, education, and numerous other topics. For pension purposes, we are interested in the individual's pension plan contributions. Unfortunately, the PSID database does not offer a full, cross-year individual time series of pension contributions, so we use the total wage earned from labor instead and assume a contributed ratio. Furthermore, income attributable to bonuses, independent businesses, secondary professional practices, comission, tips and other sources is not incorporated, due to the discontinuous and sparse nature of data. The constructed trajectories span the period from 1970 through 1992. Overall, 58,807 individuals were considered, out of which only 3,669 began working in 1970. The number of individuals starting to work each year is given in figure \ref{fig:psid} (p.\pageref{fig:psid}).

All of the wage data are adjusted to the cost of living, using the Bureau of Labor Statistics's historical CPI values. For highly-noisy volatility estimation, 3\% of the largest volatility values were omitted as outliers and 5\% of the highest salary growths were discarded as well (25,000\% growth rates etc).

The drift and diffusion coefficients of the wages model are calculated as in section \ref{sec:drift_and_diffusion_surface_interpolation}. Specifically \cite{paz},
\begin{align}
ds_i(t) = \xi s_i(t)\,dt + \eta s_i(t) dw_i(t),\quad s_i(t_{i_0}) = 1, \label{s_sde}
\end{align}
with
\begin{align}
\xi(t) \equiv -0.0328,\quad \eta(t) \equiv \sqrt{\frac16}.\label{xieta}
\end{align}
\subsection{Construction of a stochastic model of the pension fund}
In our model of the pension plan, the assets of the pension funds are invested a stock-market index, such as S\&P500. The value of the index is the market-capitalization weighted average of its components' stock prices.

We use the following definitions:
\begin{itemize}
\item $v_i(t)=$ the growth in the amount payable by the fund to member $i$ by time $t$.
\item $T_i = \{t_{i_0} < t_{i_1} < \ldots < t_{i_n}=t\}=$ the equipartition of the interval
$[t_{i_0},t]$ corresponding to the contributions of the $i$-th member to the pension fund.
\item $c_i(t) = \Lambda s_i(t)=$ the total contribution (both of employer and employee) is a
constant fraction $\Lambda$ of the salary (around $10\%$).
\item $\alpha_i=$ member's $i$ first salary (in US dollars).
\end{itemize}
We incorporate the model $Z_n(t)$ of the approximated portfolio returns to the derivation of the
equation for $v_i(t)$ by making the following observation. For every $0\leq j \leq n$, the
contributed dollar amount at time $t_{i_j}$ is given by $\alpha_i c_i(t_{ij})$, where the appreciation of  this amount is compounded from $t_{i_j}$ through $t_{i_n}$ and is given by
$Z_n(t_{i_n})/Z_n(t_{i_j})$. Therefore, the portion of the pension's total amount, attributable to $j$-th contribution, is given by
\begin{align}
\alpha_i c_i(t_{i_j}) \frac{Z_n(t)}{Z_n(t_{i_j})} \label{salary_j_contribution}
\end{align}
and the portion of the pension's total growth, attributable to the $j$-th contribution, is obtained from \eqref{salary_j_contribution} by division by $\alpha_i$. Therefore, the total growth of the pension fund, from time $t_{i0}$ to time $t$, is given by
\begin{align}
v_i(t) = \sum\limits_{\tau \in T_i}c_i(\tau)\frac{Z_n(t)}{Z_n(\tau)} = Z_n(t)\sum\limits_{\tau \in T_i}\frac{c_i(\tau)}{Z_n(\tau)}. \label{vi_discrete}
\end{align}
The continuous model for $v_i(t)$ is obtained by representing \eqref{vi_discrete} as the Riemann sum
\begin{align}
v_i(t) = \frac{Z_n(t)}{\Delta t}\sum\limits_{j=1}^{n}\frac{c_i(j\Delta t)}{Z_n(j\Delta t)}\Delta t
\end{align}
where $\Delta t = (t_{i_j}-t_{i_{j-1}})$  is the constant time elapsed between consecutive salaries. Based on the PSID database, $\Delta t = 1$ year. We write $v_i(t)$ in the integral form\\

\begin{align}
v_i(t) = Z_n(t)\left(\int\limits_{t_{i_0}}^{t}\frac{c_i(u)}{Z_n(u)}\,du\right), \label{v_i_integral}
\end{align}
and the absolute amount payable to member $i$  in dollars can now be expressed as
\begin{align}
V_i(t)=\alpha_i v_i(t).
\end{align}
We obtain the SDE for $v_i(t)$ by differentiating \eqref{v_i_integral}, and applying the chain rule
\begin{align}
dv_i(t) = dZ_n(t)\left(\int\limits_{t_{i_0}}^{t}\frac{c_i(u)}{Z_n(u)}\,du\right) + Z_n(t)\left(\frac{c_i(t)}{Z_n(t)}\,dt\right). \label{v_i_sde}
\end{align}
Substituting \eqref{Z_SDE}, $c_i(t) = \Lambda s_i(t)$, and \eqref{v_i_integral} into \eqref{v_i_sde}, we obtain the SDE
\begin{align}
dv_i(t) &= \left[\psi(t)v_i(t) + \Lambda s_i(t) \right]\,dt + \Phi(t) v_i(t)\,dW(t). \label{v_i_sde}
\end{align}
The Fokker-Planck equation for the joint probability density function $p(v,s,t)$ of the solution $\left(s_i(t),v_i(t)\right)$ of the system \eqref{s_sde}, \eqref{v_i_sde} is given by
\begin{align}
\frac{\partial p}{\partial t} = -\frac{\partial}{\partial v}\left[\left(\psi v + \Lambda s\right)p\right] -\frac{\partial}{\partial s}(\xi sp) + \frac12 \frac{\partial^2}{\partial s^2}\left( \eta^2s^2p\right) + \frac12 \frac{\partial^2}{\partial v^2}\left[\Phi^2(t)v^2p\right].\label{vi_FPE}
\end{align}
\subsection{The probability density function of the pension fund}
The probability distribution function of $V_i(t)$ is the probability that the pension payable to individual $i$ at time $t$ exceeds $y$ dollars. To compute,
\begin{align}
\Pr\left(V_i(t) > y\right) =& \Pr\left\{ v_i(t) > \frac{y}{\alpha_{t_{i_0}}} \right\} = 1 - \Pr\left\{v_i(t) \leq \frac{y}{\alpha_{t_{i_0}}}\right\},
\end{align}
we compute the joint transition probability density function $p(v,s,t \mid v_0,s_0,t_{i_0})$ of the process $v_i(t)$ and $s_i(t)$ from the FPE \eqref{vi_FPE} with the initial condition
\begin{align}
p(v,s,t_{i_0} \mid v_0, s_0, t_{i_0}) = \delta(v-v_0,s-s_0),\label{vi_FPE_ic}
\end{align}
where $v_0 = s_0 = 1$.

\subsubsection{Boundary conditions}

The stochastic process $s_i(t)$ is always positive, because it is an exponential of a Gaussian process. Furthermore, the stochastic process $v_i(t)$ is always positive, because it is a sum of a product of a lognormal and a ratio of two lognormal processes. Therefore the joint density cannot contain any mass on the boundary and thus
\begin{align}
p(v,0,t) = p(0,s,t) = 0. \label{implicit_bc_physical}
\end{align}
Because the boundaries $v=0$ and $s=0$ are unattainable by the stochastic processes, the conditions \eqref{implicit_bc_physical} are set numerically.

The distant boundaries of the grid describe the possibility of the market/salaries to get within given time to unheard of levels. Because there is no data at such levels, a zero condition for the FPE at distant boundaries $v=N_v,s=N_s$, so that the grid covers the rectangle $G=\{0<v<N_v,\,0<s<N_s\}$. The boundary $\partial G$ is a part of the model and concurs with the data. Consequently, we set
\begin{align}
p(v,s,t)\bigg|_{\partial G} = 0. \label{fpe_distant_boundary}
\end{align}

\subsubsection{The initial condition} \label{sec:ss_ic}
We approximate $\delta(v-v_0,s-s_0)$ numerically by a multivariate normal distribution with a small standard deviation and  set $p_{j,l}^{0}$ to be the multivariate Gaussian with covariance matrix $$\Sigma = \left(\begin{array}{cc} \sigma_v^2 & 0 \\ 0 & \sigma_s^2 \end{array}\right),$$
where $\sigma_v,\sigma_s \ll 1$ , and mean $\boldsymbol{\mu} = (v_0,s_0)^T = (1,1)^T$. For every $\boldsymbol{v} = (v ,s)^T \in \mathbb{R}^2$
\begin{align}
p_{j,l}^{0} =& \frac{1}{2\pi \sqrt{\det(\Sigma)}}\exp\left\{-\frac12 (\boldsymbol{v} - \boldsymbol{\mu})^T \Sigma^{-1}(\boldsymbol{v} - \boldsymbol{\mu}) \right\} = \nonumber \\
=& \frac{1}{2\pi \sigma_v\sigma_s} \exp\left\{ -\frac{(v-1)^2}{2\sigma_v^2} -\frac{(s-1)^2}{2\sigma_s^2} \right\}. \label{numerical_ic}
\end{align}
We discretize $\mathbb{R}^3$ on a $(t,v,s)$ grid with steps $(\Delta k, \Delta h , \Delta m)$ and abbreviate
\begin{align}
p(v_j,s_l,t_n) = p_{j,l}^{n} , \hspace{0.5em}\mbox{for}\ j,l,n \geq 0,
\end{align}
where $v_j = j\Delta h$ , $s_l = l\Delta m$ and $t_n = n\Delta k$.
The initial density is normalized by $\hat p_{j,l}^{0} = p_{j,l}^{0} / \int_{\mathbb{R}^2} p_{j,l}^{0}$ to insure the normalization $\int_{-\infty}^{\infty}\int_{-\infty}^{\infty}p(v,s,0)\,ds\,dv = 1$.

{The location of the zero conditions has negligible effect on the solution in the domain where it does not vanish. Extending the grid boundaries from $(N_v,N_s)$ to $(N_v',N_s')$ with $N_v' > N_v , N_s' > N_s$ introduces a change in the linear equations of the finite difference scheme that corresponds to the boundaries. Under an implicit method scheme (see below), the error infiltrates the interior of the domain by a coefficient that depend on $\Delta m^2$, $\Delta h^2$ and $\Delta k$. Therefore, if we show that
\begin{align}
\int\limits_{0}^{N_v'}\int\limits_{0}^{N_s'}p_{j,l}^{0}dsdv - \int\limits_{0}^{N_v}\int\limits_{0}^{N_s}p_{j,l}^{0}dsdv \label{numerical_boundary_extension}
\end{align}
is negligible, then by the above argument, we conclude that the change in the solution is negligible. There are several numerical methods for approximating the normal CDF \cite{normal_cdf_approx_1}, \cite{normal_cdf_approx_2}, \cite{normal_cdf_approx_3}. We compute \eqref{numerical_boundary_extension} numerically for grids  $18\times5$  extended to $36\times10$. The numerical analysis of the computation is given in Appendix 1 below.
The numerical precision is summarized in  table \ref{tab:G}.
\begin{table}
\begin{center}
\begin{tabular}{|c|c|c|c|c|c|c|c|c|}
\hline
$\Delta h$ & $\Delta m$ & $N_v$ & $N_s$ & $N_v'$ & $N_s'$ & $N_v\Delta h$ & $N_s\Delta m$ & \eqref{numerical_boundary_extension} \\
\hline
0.025 & 0.2 & 720 & 25 & 1440 & 50 & 18 &5 & $1.0805 \cdot 10^{-172}$ \\ \hline
\end{tabular}
\caption{\small Grid size, domain size, and error.\label{tab:G}}
\end{center}
\end{table}
}

\subsubsection{Results}
A numerical solution to the initial and boundary value problem for the FPE is constructed by the
finite difference method (FDM). We use the stable implicit BTCS (First Order Backward Time Central
Space) \cite{crank} method to approximate $p(v,s,t)$ in $G$ for $t>t_0$.
The probability that the pension fund will be of size $y$ equals the probability that the fund's
growth will equal the proportion of $y$ and the initial salary $\alpha$. This ratio is the
dimensionless parameter of the problem. In the table below, we show probabilities for different
ratios with the assumption of a 10\% salary contribution. We say that a target pension $y$ , with
initial salary $\alpha$  and $t$ years of savings, has an implied annual return $r$, if
$\sum_{i=1}^{t}\alpha\left(1+r\right)^i = y$.
\renewcommand{\arraystretch}{1.75}
\begin{table}
\begin{center}
\begin{tabular}{l|c|c|r}
\hline
$\frac{\mbox{Target Pension Size}}{\mbox{Initial Salary}}$ & Saving Period  & Implied Annual Return & Probability \\
\hline\hline
3.11      & 25 Years & 1.64\% & 65.40\%  \\ \hline
3.33      & 25 Years & 2.15\% & 54.40\%  \\ \hline
3.55      & 25 Years & 2.61\% & 45.17\%  \\ \hline
4.00      & 25 Years & 3.60\% & 28.27\%  \\ \hline
%4.16      & 25 Years & 3.73\% & 23.70\%  \\ \hline
4.44      & 25 Years & 4.17\% & 16.16\%  \\ \hline
5.00      & 25 Years & 4.98\% & 7.37\%  \\ \hline
5.83      & 25 Years & 6.02\% & 1.72\%  \\ \hline
6.67      & 25 Years & 6.90\% & 0.34\%  \\ \hline\hline
\end{tabular}
\caption{Pension size probabilities for 25 years of savings.}
\end{center}
\end{table}
\begin{table}
\begin{center}
\begin{tabular}{l|c|c|r}
\hline
$\frac{\mbox{Target Pension Size}}{\mbox{Initial Salary}}$ & Saving Period  & Implied Annual Return & Probability \\
\hline\hline
5.00      & 40 Years & 1.05\% & 59.38\%  \\ \hline
6.50      & 40 Years & 2.23\% & 54.51\%  \\ \hline
7.00      & 40 Years & 2.55\% & 49.17\%  \\ \hline
7.50      & 40 Years & 2.85\% & 41.77\%  \\ \hline
9.50      & 40 Years & 3.83\% & 21.69\%  \\ \hline
11.00      & 40 Years & 4.43\% & 14.86\%  \\ \hline
15.00      & 40 Years & 5.65\% & 1.07\%  \\ \hline\hline
\end{tabular}
\caption{Pension size probabilities for 40 years of savings.}
\end{center}
\end{table}
\renewcommand{\arraystretch}{1}

\begin{figure}[ht!]
\begin{center}
\begin{tabular}{c}
\includegraphics[width=0.4\textwidth,height=0.3\textheight]{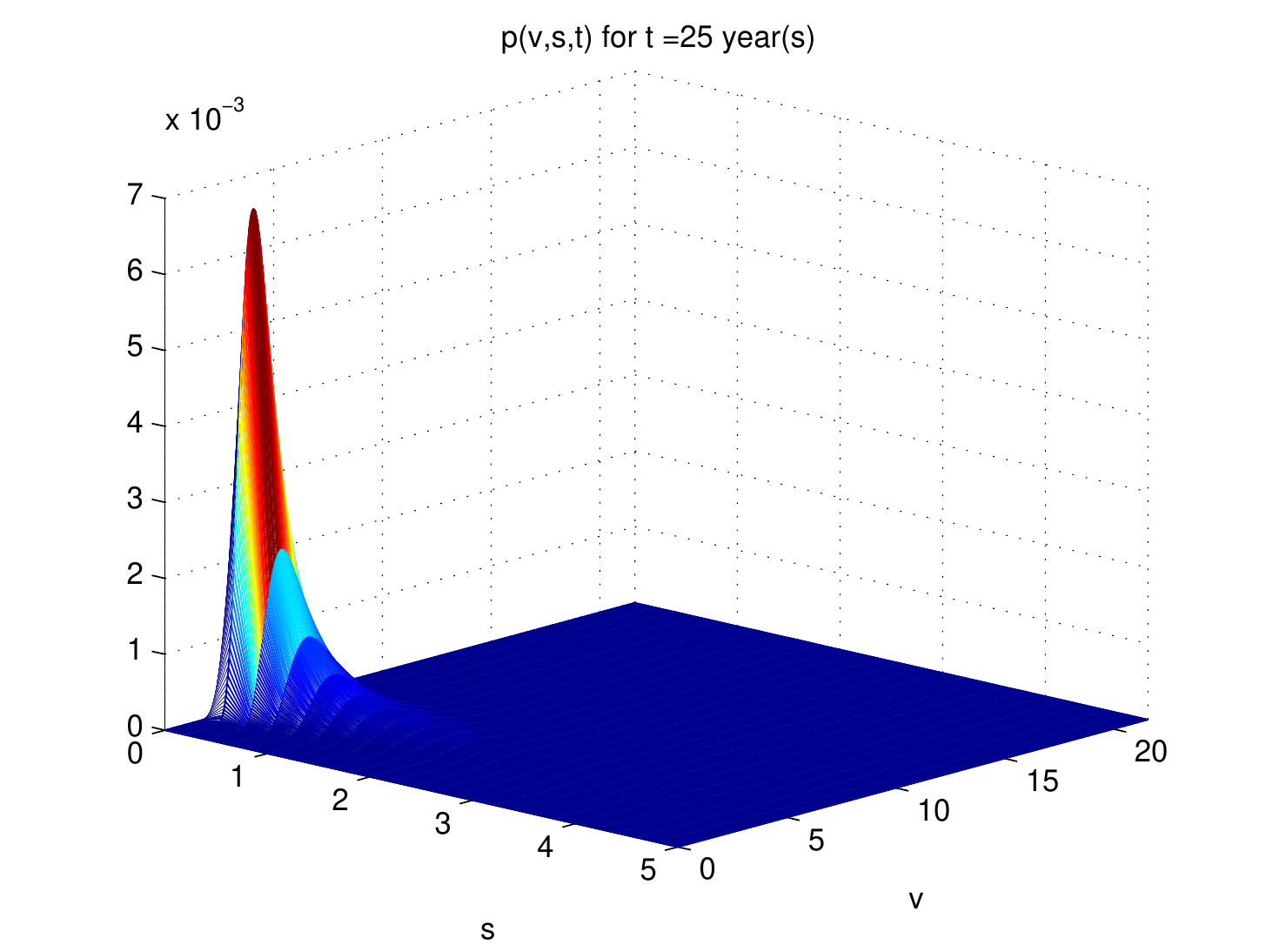}
\includegraphics[width=0.4\textwidth,height=0.3\textheight]{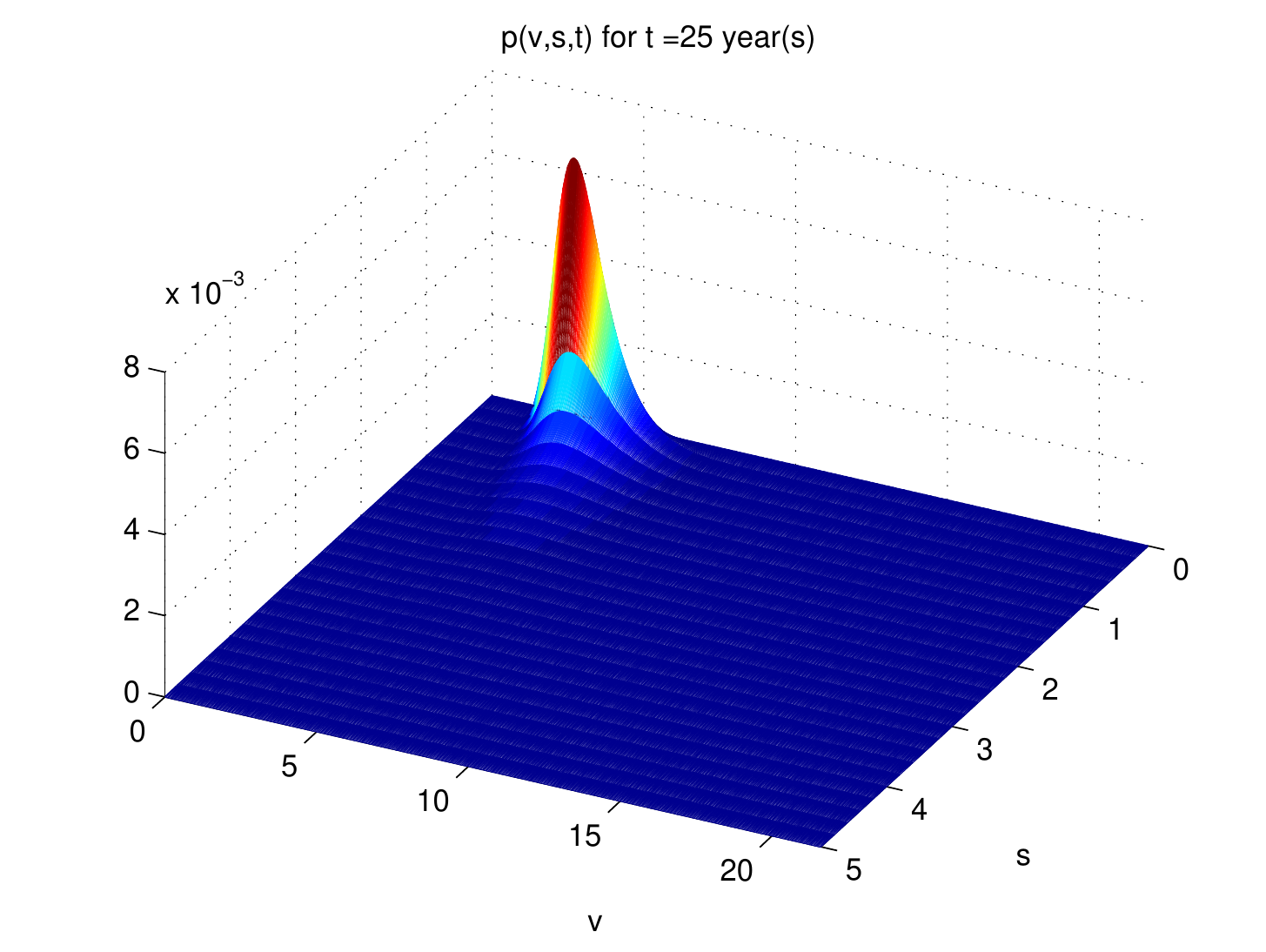}
\end{tabular}
\end{center}
\caption{\small The joint pdf $p(v,s,t)$ for 25 years. The grid on the $v$-axis has 720 points, spaced 0.03 apart, that on the $s$-axis has 25 points, spaced 0.2 apart. $\Delta t$ = 0.1 \label{fig:pdf25_1} }
\end{figure}

\begin{figure}[ht!]
\begin{center}
\begin{tabular}{c}
\includegraphics[width=0.4\textwidth,height=0.3\textheight]{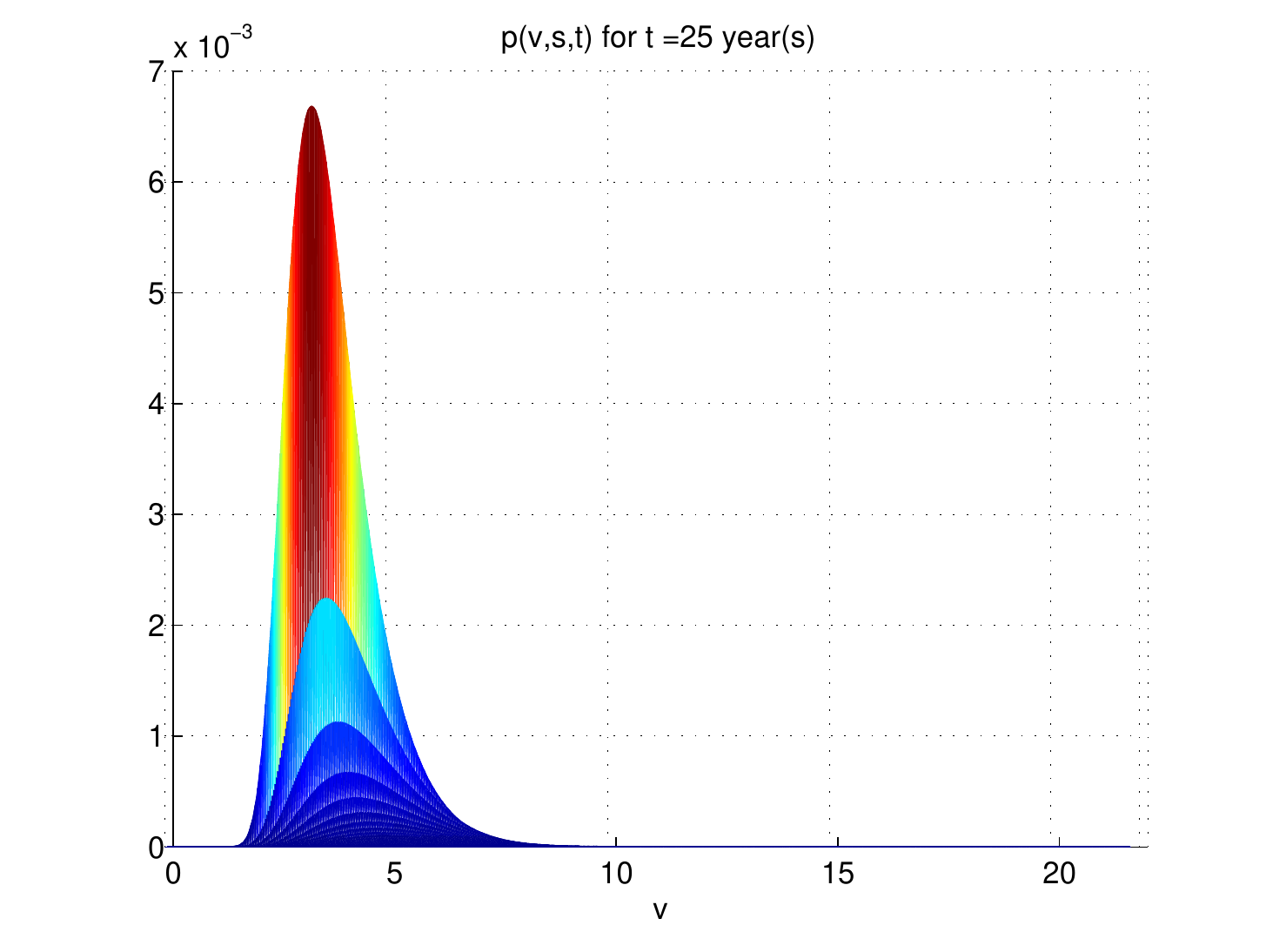}
\includegraphics[width=0.4\textwidth,height=0.3\textheight]{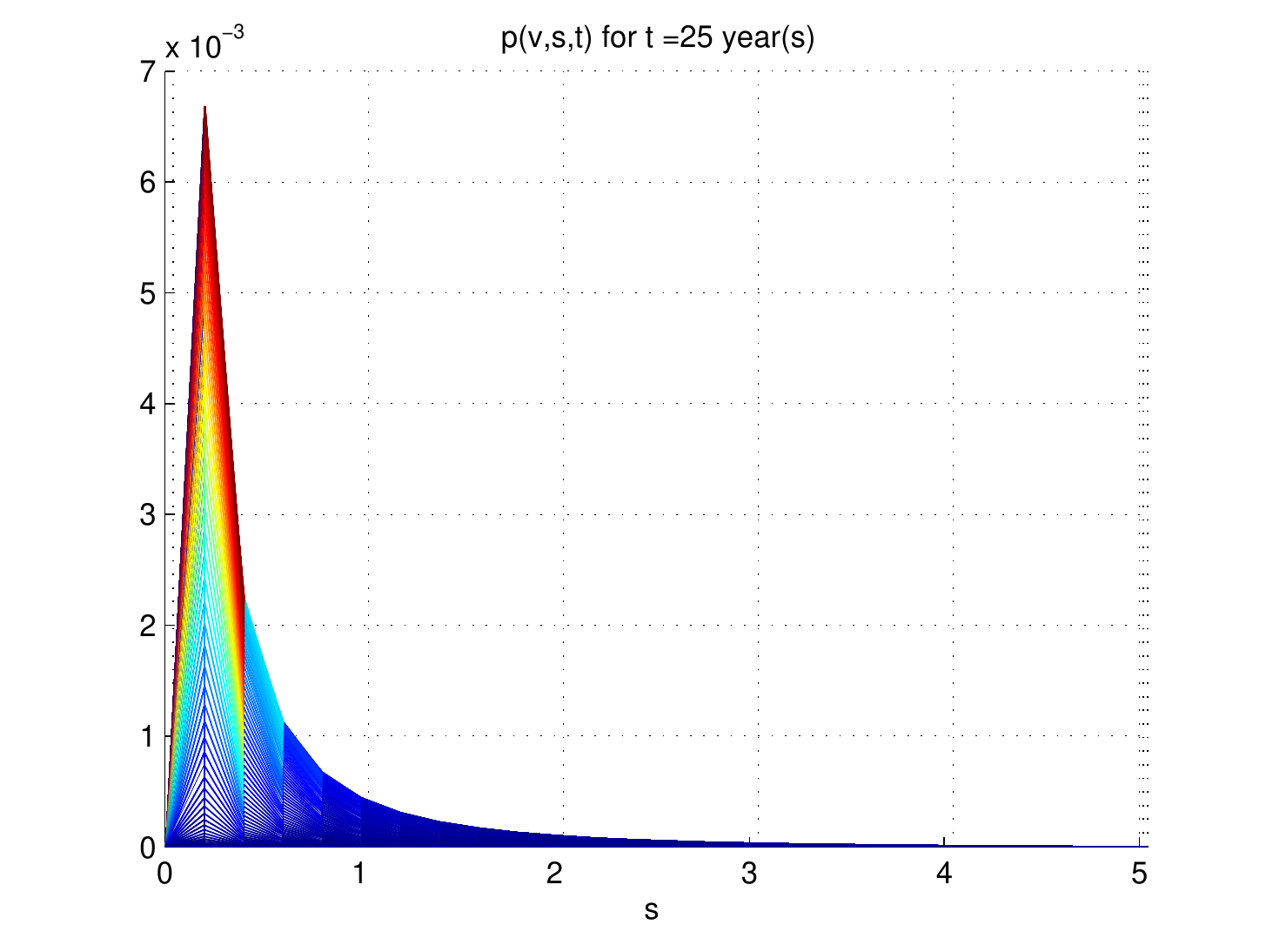}
\end{tabular}
\end{center}
\caption{\small The joint pdf $p(v,s,t)$ for 25 years. {\bf (Left) } $p(v,s,t)$ projected on the $v$-axis. {\bf (Right) } $p(v,s,t)$ projected on the $s$-axis }
\label{fig:pdf25_projections}
\end{figure}

The three-dimensional graphs of the joint density function $p(v,s,t)$ are given in figure \ref{fig:pdf25_1} for $t=25$
and in figure \ref{fig:pdf25_projections} for $t=50$
\begin{figure}[ht!]
\begin{center}
                \includegraphics[width=0.4\textwidth,height=0.3\textheight]{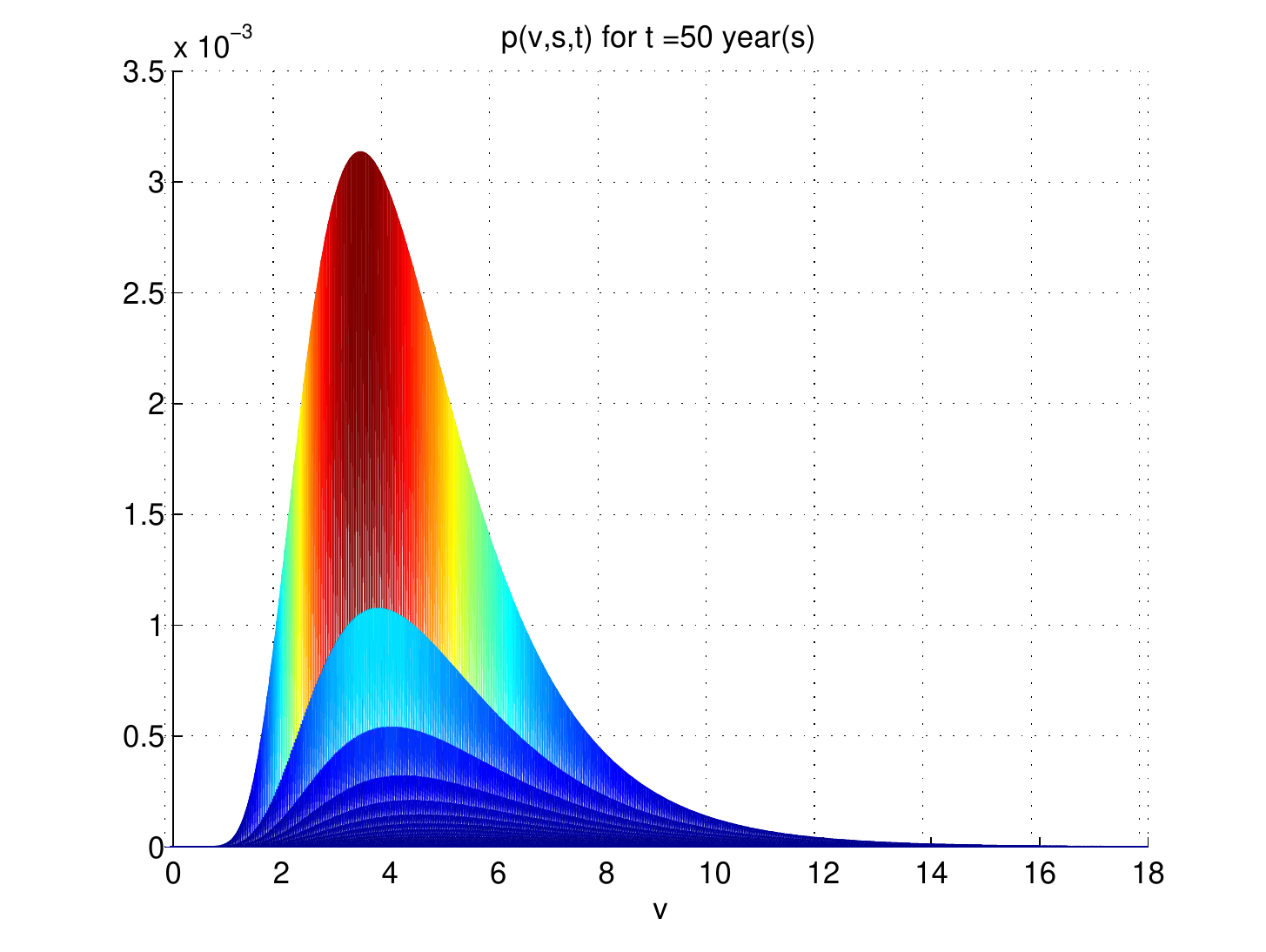}
               \includegraphics[width=0.4\textwidth,height=0.3\textheight]{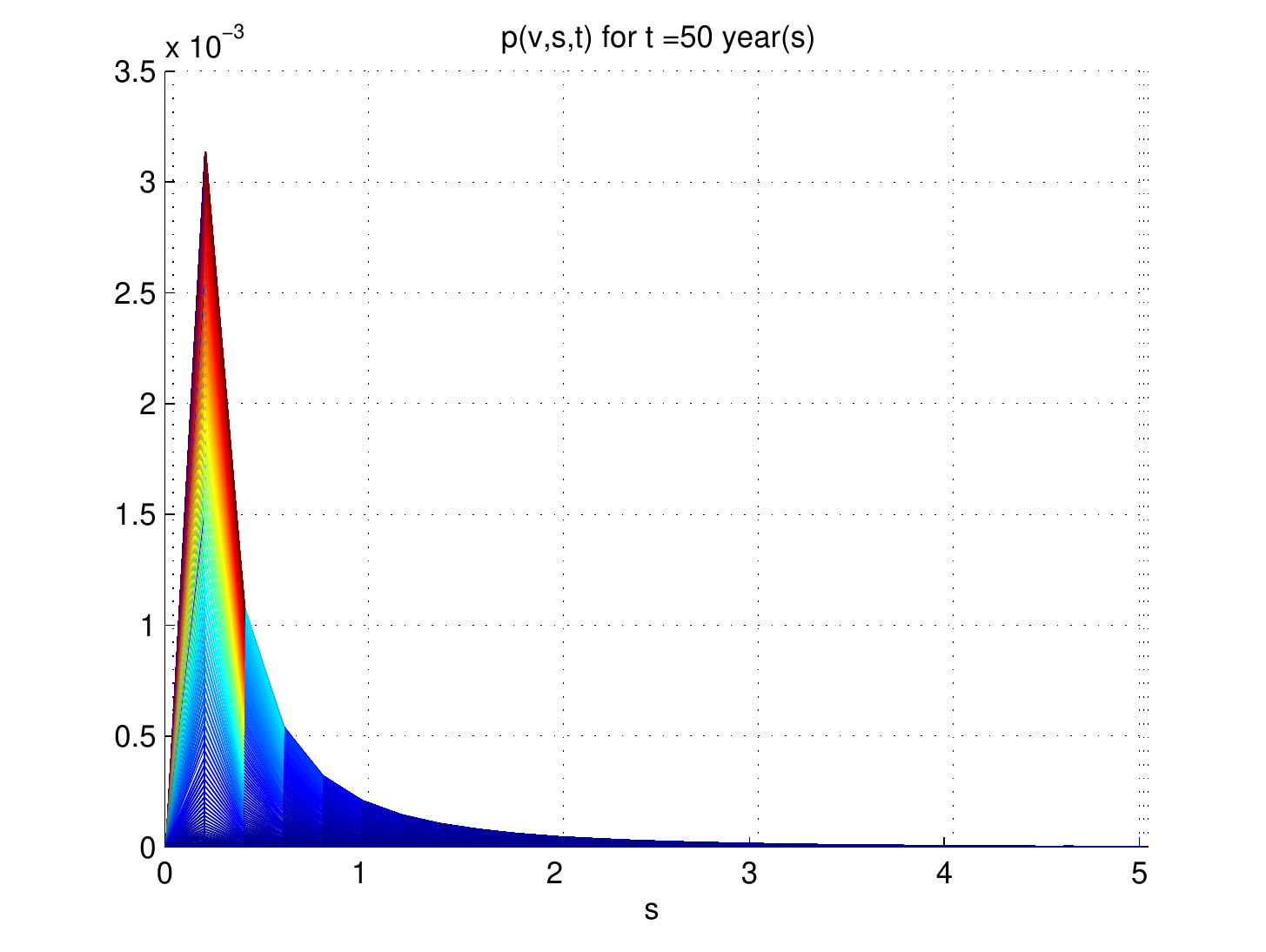}
        \caption{\small The joint pdf $p(v,s,t)$ for 50 years. {\bf (Left) } $p(v,s,t)$ projected on the $v$-axis. {\bf (Right) } $p(v,s,t)$ projected on the $s$-axis. }
\label{fig:pdf50_projections}
\end{center}
\end{figure}
 (pp.\pageref{fig:pdf25_1}).
\begin{figure}[ht!]
\begin{center}
\begin{tabular}{c}
\includegraphics[width=0.4\textwidth,height=0.3\textheight]{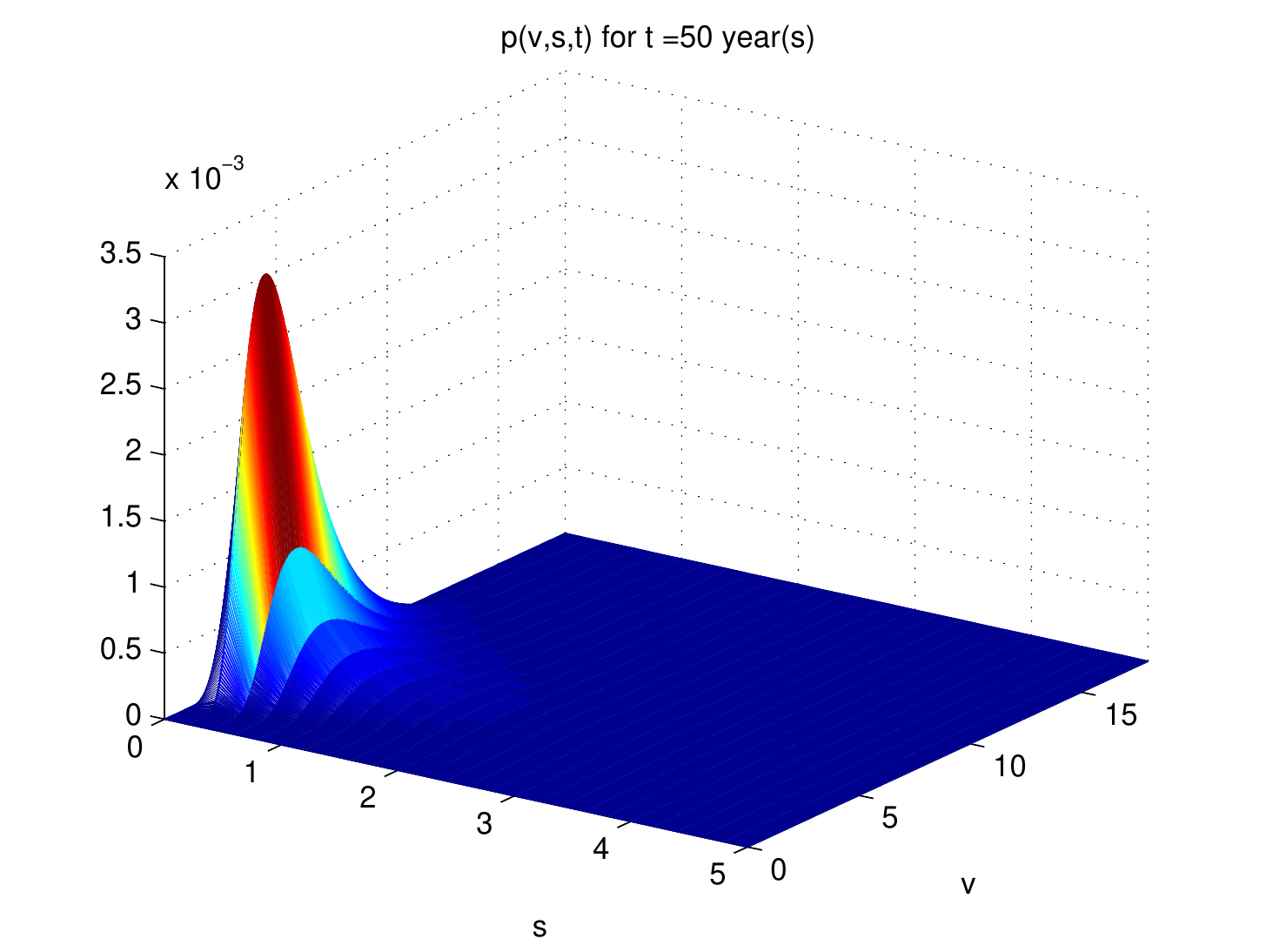}
\includegraphics[width=0.4\textwidth,height=0.3\textheight]{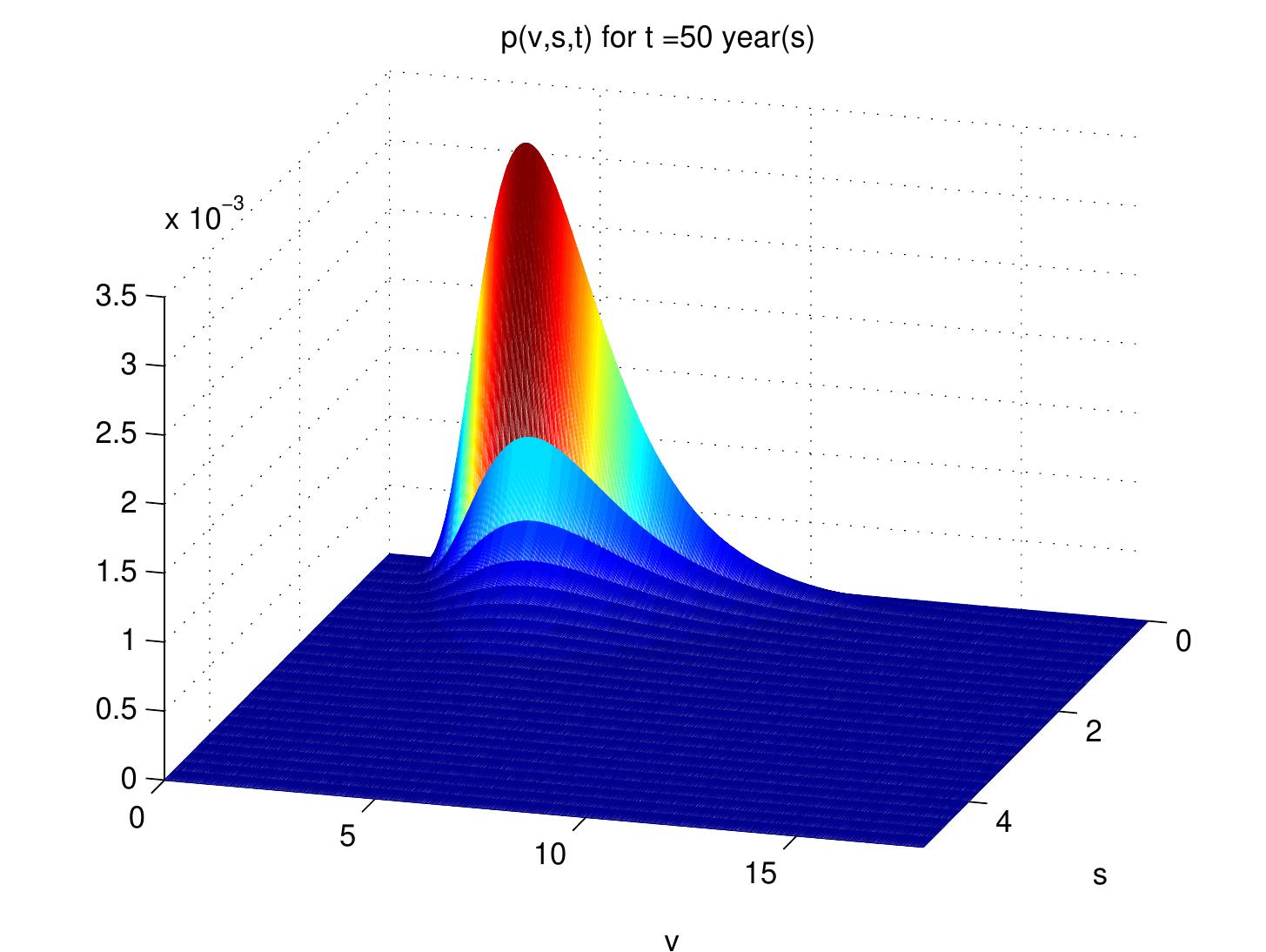}
\end{tabular}
\end{center}
\caption{\small The joint pdf $p(v,s,t)$ for 50 years. The grid on $v$-axis has 720 points, spaced 0.03 apart, that on the $s$-axis has 25 points, spaced 0.2 apart. $\Delta t$ = 0.1  }
\label{fig:pdf50_1}
\end{figure}
\subsection{Stochastic model for pension consumption}
We consider the pension consumption process, $\tilde V_i(t)$, which is the remaining dollar amount
individual $i$ has after being retired for $t - t_{0}$ years, where $t_{0}$ is the time of
retirement and $\beta_i$ is a constant annual dollar consumption rate. Assume $T_i = \{t_{i_0} <
t_{i_1} < \ldots < t_{i_n}=t\}$ are the equispaced years that member $i$ has been retired. The
initial condition for $V_i(t)$ is given by $\tilde V_i(t_0) = V_r$, where $V_0$ is the pension
accumulated during the working life of individual $i$. Upon retirement the salary stops. We
incorporate the model of the approximated portfolio returns to derive the equation for $\tilde
V_i(t)$ by shifting the initial condition of $Z_n$, thereby obtaining a new process $\tilde Z_n(t)$,
 \begin{align}
d \tilde Z_n(t) = \psi(t)\tilde Z_n(t)\,dt + \Phi(t)\tilde Z_n(t)\,dW(t),\hspace{0.5em}\mbox{for} \ t>t_{i_0},\quad \tilde Z_n(t_{i_0}) = Z_r \label{Z_tilde_SDE}
\end{align}
and making the following observation; for every $0 < j \leq n$, the value of $\tilde V_i(t_{i_j})$
grew by $\tilde Z_n(t_{i_j})/\tilde Z_n(t_{i_{j-1}})$, relative to the previous year $\tilde
V_i(t_{i_{j-1}})$, while $\beta_i$ dollars were consumed. Therefore, the value of $\tilde
V_i(t_{i_j})$ is given by the recursion relation
\begin{align*}
\tilde V_i(t_{i_j}) = \tilde V_i(t_{i_{j-1}})\frac{\tilde Z_n(t_{i_j})}{\tilde Z_n(t_{i_{j-1}})} - \beta_i, \hspace{0.5em}\mbox{for}\ 0<j\leq n,\quad\tilde V_i(t_{i_0}) =V_r
 \end{align*}
and in its closed form,
\begin{align}
\tilde V_i(t) = V_r\frac{\tilde Z_n(t)}{Z_r} - \beta \sum\limits_{j=1}^{n}\frac{\tilde Z_n(t)}{\tilde Z_n(t_{i_j})} = \tilde Z_n(t)\left(\frac{V_r}{Z_r} - \beta \sum\limits_{j=1}^{n}\frac{1}{\tilde Z_n(t_{i_j})} \right). \label{V_i_tilde_discrete}
\end{align}
The continuous model for $\tilde V_i(t)$ is obtained by representing \eqref{V_i_tilde_discrete} as the Riemann sum
\begin{align}
\tilde V_i(t) = \tilde Z_n(t)\left(\frac{V_r}{Z_r} - \frac{\beta_i}{\Delta t}\sum\limits_{j=1}^{n}\frac{\Delta t}{\tilde Z_n(t_{i_0} + j\Delta t)} \right),
\end{align}
where $\Delta t = (t_{i_j}-t_{i_{j-1}})=1$ is the constant time elapsed between consecutive time periods (time is measured in years). The process $\tilde V_i(t)$ is approximated by the integral
\begin{align}
\tilde V_i(t) = \tilde Z_n(t)\left(\frac{V_r}{Z_r} - \beta_i \int\limits_{t_{i_0}}^{t}\frac{du}{\tilde Z_n(u)}\right). \label{V_i_tilde_continuous}
\end{align}
We differentiate \eqref{V_i_tilde_continuous} to obtain the SDE
\begin{align}
d\tilde V_i(t) =& d\tilde Z_n(t)\left(\frac{V_r}{Z_r} - \beta_i \int\limits_{t_{i_0}}^{t}\frac{du}{\tilde Z_n(u)}\right) + \tilde Z_n(t)\hspace{0.2em}d\left(\frac{V_r}{Z_r} - \beta_i\int\limits_{t_{i_0}}^{t}\frac{du}{\tilde Z_n(u)}\right) \nonumber \\
=&d\tilde Z_n(t)\left(\frac{V_r}{Z_r} - \beta_i \int\limits_{t_{i_0}}^{t}\frac{du}{\tilde Z_n(u)}\right) + \tilde Z_n(t)\left(-\beta_i\frac{dt}{Z_n(t)}\right).\label{dVt}
\end{align}
Now, using \eqref{Z_tilde_SDE} and \eqref{V_i_tilde_continuous} in \eqref{dVt}, we obtain the nonhomogeneous linear SDE
\begin{align}
d\tilde V_i(t) =\left[\psi(t)\tilde V_i(t) - \beta_i \right]\,dt + \Phi(t)\tilde V_i(t)\,dW(t),\quad \tilde V_i(t_{i_0}) = V_r. \label{V_tilde_SDE}
\end{align}
The solution of \eqref{V_tilde_SDE} is given by \eqref{nonhomo_linear_sde_solution}, which reduces to
\begin{align}
\tilde V_i(t) =H(t)\left[1 -\beta_i\int\limits_{t_{i_0}}^{t}\frac{ds}{H(s)}\right],\hspace{0.5em} \mbox{for}\ t>t_{i_0},\quad
\tilde V_i(t_{i_0}) = V_r, \label{V_i_tilde_explicit_solution}
\end{align}
where
\begin{align*}
H(t) = V_r\exp\left\{\int\limits_{t_{i_0}}^{t}\left[\psi(s) - \frac12 \Phi^2(s)\right]ds + \int\limits_{t_{i_0}}^{t}\Phi(s)dW(s)\right\}.
\end{align*}
We conclude from \eqref{V_i_tilde_explicit_solution} that $\tilde V_i(t) = 0$ for $t>t_{i_0}$, so that
$$\int\limits_{t_{i_0}}^{t}\frac{ds}{H(s)} = \frac{1}{\beta_i}.$$
The Fokker-Planck equation for the pdf of the solution of \eqref{V_tilde_SDE} is given by
\begin{align}
\frac{\partial \tilde p}{\partial t} = -\frac{\partial}{\partial\tilde v}\left[\left(\psi\tilde v - \beta_i\right)\tilde p\right] + \frac12\Phi^2(t) \frac{\partial^2}{\partial \tilde v^2}\left( \tilde v^2\tilde p \right) \label{consumption_fpe}
\end{align}
for $t>t_{i_0}, \tilde v>0$, where $\tilde p = \tilde p(\tilde v,t \mid V_r , t_{i_0})$ is the transition probability density function of $\tilde V_i(t)$ with the initial condition
\begin{align}
\lim_{t \to t_{i_0}}\tilde p\left(\tilde v, t \mid V_r, t_{i_0}\right) = \delta \left(\tilde v - V_r\right). \label{consumption_ic}
\end{align}

\subsubsection{Survival probability of the consumption process}
The consumption process $\tilde V_i(t)$ ends when it hits zero for the first time
$$\tau  = \inf \left\{t>t_{0} \mid \tilde V_i(t) = 0 \right\}$$
and its survival probability is defined as
\begin{align}
S(t \hspace{0.2em} | \hspace{0.2em} V_r ,t_0) &= \Pr\left(\tau > t \mid V_r ,t_0 \right) = \int\limits_{G\setminus\partial G}^{\infty}\tilde p\left(\tilde v,t \mid V_r ,t_0\right)d\tilde v, \label{survival_prob}
\end{align}
where the pdf of $\tilde V(t)$ is the solution for the FPE \eqref{consumption_fpe} for $\tilde v > 0$ with the initial condition and boundary conditions
\begin{align*}
\tilde p(\tilde v, t_0 \mid V_r,t_0) = \delta (\tilde v - V_r), \quad \tilde p(0,t \mid V_r, t_0) \hspace{0.5em}\mbox{for}\ t > t_0. \label{bc_tilde_fpe}
\end{align*}
Changing variables in \eqref{consumption_fpe} to $\tilde v = V_r x$, we obtain
\begin{align}
\frac{\partial q}{\partial t} = -\frac{\partial}{\partial x}\left[\left(\psi x - \frac{\beta_i}{V_r}\right) q\right] + \frac12\Phi^2(t) \frac{\partial^2}{\partial x^2}\left( x^2q \right)
\end{align}
for $t>t_{i_0}, x>0$, where $q(x,t)$ is the pdf of $\tilde V_i(t)/V_r$, with the initial condition and boundary conditions
$$\tilde q\left(x, t_{i_0} \mid x_0 , t_{i_0}\right) = \delta \left(x - 1\right), \quad \tilde q(0,t \mid x_0, t_{i_0}) = 0.$$

The Internal Rate of Return (IRR) of a retirement period of $n$ years is the return $r$ that satisfies
$$\sum\limits_{i=1}^{t}\frac{\beta_i}{(1+r)^i}=V_r.$$
The results are summarized in the tables below.
\begin{table}
\begin{center}
\begin{tabular}{|l|l||c|c|}
\hline
Initial Pension   & Retirement     & IRR            & Survival  \\
$\overline{\mbox{Yearly Consumption}}$ & Period & & Probability \\
\hline\hline
7.5 Years                  & 8 Years              & 1.45\%         &    48.73\%     \\ \cline{3-4}
                  & 9 Years              & 3.81\%         &    29.04\%     \\ \cline{3-4}
                  & 10 Years             & 5.6\%          &    20.46\%     \\ \cline{3-4}
                  & 11 Years             & 6.99\%         &    14.74\%      \\
\hline
10 Years                  & 10 Years             & 0.00\%         &    79.78\%     \\ \cline{3-4}
                  & 11 Years             & 1.62\%         &    54.01\%     \\ \cline{3-4}
                  & 12 Years             & 2.92\%         &    31.12\%     \\ \cline{3-4}
                  & 13 Years             & 3.97\%         &    20.6\%     \\ \cline{3-4}
                  & 14 Years             & 4.84\%         &    14.75\%      \\ \cline{3-4}
                  & 15 Years             & 5.55\%         &    10.79\%      \\
\hline
12 Years                  & 13 Years             & 1.16\%         &    70.79\%     \\ \cline{3-4}
                  & 14 Years             & 2.12\%         &    48.21\%     \\ \cline{3-4}
                  & 15 Years             & 2.92\%         &    29.22\%     \\ \cline{3-4}
                  & 16 Years             & 3.60\%         &    18.53\%     \\ \cline{3-4}
                  & 17 Years             & 4.17\%         &    12.7\%      \\ \cline{3-4}
                  & 18 Years             & 4.66\%         &    9.11\%      \\
\hline
\end{tabular}
\caption{\small Survival probabilities for consumption periods of 7.5, 10, and 12 years of uninvested pension.}
\end{center}
\end{table}
\begin{table}
\begin{center}
\begin{tabular}{|l|l||c|c|}
\hline
Initial Pension   & Retirement     & IRR            & Survival  \\
$\overline{\mbox{Yearly Consumption}}$ & Period & & Probability \\
\hline\hline
12.5 Years                  & 13 Years             & 0.56\%         &    82.36\%     \\ \cline{3-4}
                  & 14 Years             & 1.54\%         &    64.46\%     \\ \cline{3-4}
                  & 15 Years             & 2.37\%         &    42.61\%     \\ \cline{3-4}
                  & 16 Years             & 3.06\%         &    26.14\%     \\ \cline{3-4}
                  & 17 Years             & 3.65\%         &    16.68\%     \\ \cline{3-4}
                  & 18 Years             & 4.15\%         &    11.4\%     \\ \cline{3-4}
                  & 19 Years             & 4.58\%         &    8.12\%      \\ \cline{3-4}
                  & 20 Years             & 4.96\%         &    5.84\%      \\
\hline
15 Years                    & 15 Years             & 0.00\%         &    93.17\%     \\ \cline{3-4}
                    & 20 Years             & 2.91\%         &    28.93\%     \\ \cline{3-4}
                    & 25 Years             & 4.38\%         &    3.48\%     \\ \cline{3-4}
                    & 30 Years             & 5.21\%         &    0.43\%     \\
\hline
16.25 Years                  & 20 Years             & 2.06\%         &    60.94\%     \\ \cline{3-4}
                  & 25 Years             & 3.63\%         &    9.61\%     \\ \cline{3-4}
                  & 30 Years             & 4.52\%         &    1.08\%     \\ \cline{3-4}
                  & 35 Years             & 5.06\%         &    0.09\%     \\
\hline
\end{tabular}
\caption{\small Survival probabilities for consumption periods of 12.5, 15, and 16.25 years of uninvested pension.}
\end{center}
\end{table}
\subsubsection{Mean first passage time of the consumption process}
The mean first passage Time (MFPT), is given by \cite{schuss}
\begin{align}
\eE[\tau \mid V_r,  \tau > t_0] = \int\limits_{t_0}^{\infty}\Pr\left\{ \tau > t \mid V_r , t_0 \right\}\,dt = \int\limits_{t_0}^{\infty}\int\limits_{0}^{\infty}\tilde p\left(\tilde v,t \mid V_r, t_0\right)d\tilde v\,dt.
\end{align}
The results are summarized in tables below.
\begin{table}
\begin{center}
    \begin{tabular}{|l|r|}
    \hline
    Pension/Consumption                                           & MFPT          \\ \hline
    7.50 Years                                                    & 8.27 Years    \\ \hline
    10.00 Years                                                   & 11.29 Years   \\ \hline
    12.00 Years                                                   & 13.86 Years    \\ \hline
    12.50 Years                                                   & 14.53 Years   \\ \hline
    15.00 Years                                                   & 18.16 Years    \\ \hline
    16.25 Years                                                   & 20.15 Years   \\ \hline
  \end{tabular}
  \caption{Mean first passage times for different consumption rates of uninvested pensions.}
\end{center}
\end{table}
\subsubsection{The probability that the pension survives the pensioner}
\begin{figure}[ht!]
\begin{center}
	  \includegraphics[width=0.55\textwidth,height=0.35\textheight]{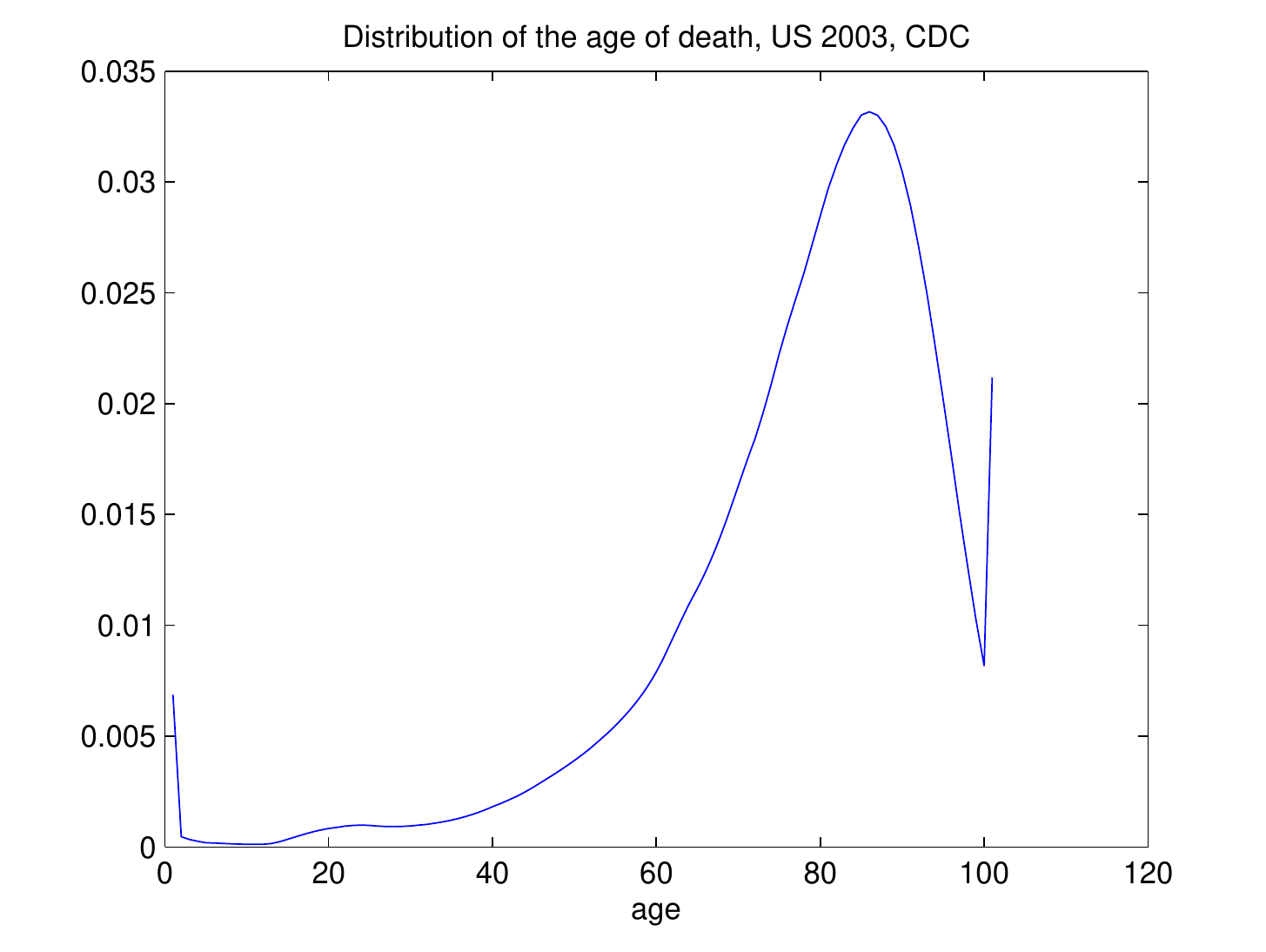}
	  \caption{\small {The pdf of the age of death, US 2003}}
	\label{fig:death_pdf}
\end{center}
\end{figure}

We assume that the time of death of a given pensioner is a random variable $T$, with pdf $f_T(t)$. The probability that the pension survives the pensioner, or the probability that the pensioner dies before he/she runs out of money, is given by
\begin{align}
\Pr\left\{\tau > T \right\} = \int\limits_{t_0}^{\infty}\Pr\{\tau > T \mid T = t\}f_T(t \mid t_0)\,dt = \int\limits_{t_0}^{\infty}S(t)f_T(t \mid t_0 )\,dt.
\end{align}
The distribution of life expectancy is taken from US Department of Health and Human Services (HHS), Centers for Disease Control and Prevention (CDC), and is based on US population \cite{cdc}. See figure \ref{fig:death_pdf}, table \ref{table:death_pdf_table} (pp.\pageref{table:death_pdf_table}--\pageref{fig:death_pdf}).
\begin{table}
\begin{center}
    \begin{tabular}{|l|c|c|c|c|c|c|}
    \hline
    Age & Probability of & Number & Number  & Person-years & Total number & Expec.   \\
    & dying between & surviving & dying  ages  & lived between  & of person-years & of life \\
    & ages $x$ to $x+1$ & to age $x$ & between ages & ages $x$ to $x+1$ & lived above & at age $x$ \\
    &               &          &  $x$ to $x+1$        &               & age $x$           &            \\
    \hline
68-69	 & 	0.019320	 & 	78,705	 & 	1521	 & 	77,944	 & 	1,275,953	 & 	16.2\\
69-70	 & 	0.021108	 & 	77,184	 & 	1629	 & 	76,369	 & 	1,198,008	 & 	15.5\\
70-71	 & 	0.022950	 & 	75,555	 & 	1734	 & 	74,688	 & 	1,121,639	 & 	14.8\\
71-72	 & 	0.024904	 & 	73,821	 & 	1838	 & 	72,902	 & 	1,046,951	 & 	14.2\\
72-73	 & 	0.027151	 & 	71,982	 & 	1954	 & 	71,005	 & 	974,050	 & 	13.5\\
73-74	 & 	0.029784	 & 	70,028	 & 	2086	 & 	68,985	 & 	903,044	 & 	12.9\\
74-75	 & 	0.032753	 & 	67,942	 & 	2225	 & 	66,830	 & 	834,059	 & 	12.3\\
75-76	 & 	0.035831	 & 	65,717	 & 	2355	 & 	64,540	 & 	767,230	 & 	11.7\\
76-77	 & 	0.038987	 & 	63,362	 & 	2470	 & 	62,127	 & 	702,690	 & 	11.1\\
77-78	 & 	0.042503	 & 	60,892	 & 	2588	 & 	59,598	 & 	640,563	 & 	10.5\\
78-79	 & 	0.046557	 & 	58,304	 & 	2714	 & 	56,947	 & 	580,965	 & 	10.0\\
79-80	 & 	0.051200	 & 	55,589	 & 	2846	 & 	54,166	 & 	524,019	 & 	9.4\\
80-81	 & 	0.056335	 & 	52,743	 & 	2971	 & 	51,258	 & 	469,853	 & 	8.9\\
81-82	 & 	0.061837	 & 	49,772	 & 	3078	 & 	48,233	 & 	418,595	 & 	8.4\\
82-83	 & 	0.067856	 & 	46,694	 & 	3168	 & 	45,110	 & 	370,362	 & 	7.9\\
83-84	 & 	0.074504	 & 	43,526	 & 	3243	 & 	41,904	 & 	325,252	 & 	7.5\\
84-85	 & 	0.081975	 & 	40,283	 & 	3302	 & 	38,632	 & 	283,348	 & 	7.0\\
85-86	 & 	0.089682	 & 	36,981	 & 	3317	 & 	35,322	 & 	244,716	 & 	6.6\\
86-87	 & 	0.098031	 & 	33,664	 & 	3300	 & 	32,014	 & 	209,394	 & 	6.2\\
87-88	 & 	0.107059	 & 	30,364	 & 	3251	 & 	28,739	 & 	177,380	 & 	5.8\\
88-89	 & 	0.116804	 & 	27,113	 & 	3167	 & 	25,530	 & 	148,641	 & 	5.5\\
89-90	 & 	0.127300	 & 	23,946	 & 	3048	 & 	22,422	 & 	123,111	 & 	5.1\\
90-91	 & 	0.138581	 & 	20,898	 & 	2896	 & 	19,450	 & 	100,689	 & 	4.8\\
91-92	 & 	0.150676	 & 	18,002	 & 	2712	 & 	16,646	 & 	81,239	 & 	4.5\\
92-93	 & 	0.163611	 & 	15,289	 & 	2502	 & 	14,039	 & 	64,594	 & 	4.2\\
93-94	 & 	0.177408	 & 	12,788	 & 	2269	 & 	11,654	 & 	50,555	 & 	4.0\\
94-95	 & 	0.192080	 & 	10,519	 & 	2021	 & 	9,509	 & 	38,901	 & 	3.7\\
95-96	 & 	0.207636	 & 	8,499	 & 	1765	 & 	7,616	 & 	29,392	 & 	3.5\\
96-97	 & 	0.224075	 & 	6,734	 & 	1509	 & 	5,980	 & 	21,776	 & 	3.2\\
97-98	 & 	0.241387	 & 	5,225	 & 	1261	 & 	4,594	 & 	15,796	 & 	3.0\\
98-99	 & 	0.259552	 & 	3,964	 & 	1029	 & 	3,449	 & 	11,202	 & 	2.8\\
99-100	 & 	0.278539	 & 	2,935	 & 	818	 & 	2,526	 & 	7,752	 & 	2.6\\
100+	 & 	1.00000	 & 	2,118	 & 	2118	 & 	5,226	 & 	5,226	 & 	2.5 \\ \hline
    \end{tabular}
\caption{\small The distribution of life span in the USA, 2003 (CDC)}
\label{table:death_pdf_table}
\end{center}
\end{table}
\begin{table}{}
\begin{center}
    \begin{tabular}{|l|c|c|c|c|c|c|}
    \hline
    Age & Probability of & Number & Number  & Person-years & Total number & Expec.   \\
    & dying between & surviving & dying  ages  & lived between  & of person-years & of life \\
    & ages $x$ to $x+1$ & to age $x$ & between ages & ages $x$ to $x+1$ & lived above & at age $x$ \\
    &               &          &  $x$ to $x+1$        &               & age $x$           &            \\
    \hline
    0-1	 & 	0.006865	 & 	100,000	 & 	687	 & 	99,394	 & 	7,743,016	 & 	77.4\\
1-2	 & 	0.000469	 & 	99,313	 & 	47	 & 	99,290	 & 	7,643,622	 & 	77.0\\
2-3	 & 	0.000337	 & 	99,267	 & 	33	 & 	99,250	 & 	7,544,332	 & 	76.0\\
3-4	 & 	0.000254	 & 	99,233	 & 	25	 & 	99,221	 & 	7,445,082	 & 	75.0\\
4-5	 & 	0.000194	 & 	99,208	 & 	19	 & 	99,199	 & 	7,345,861	 & 	74.0\\
5-6	 & 	0.000177	 & 	99,189	 & 	18	 & 	99,180	 & 	7,246,663	 & 	73.1\\
6-7	 & 	0.000160	 & 	99,171	 & 	16	 & 	99,163	 & 	7,147,482	 & 	72.1\\
7-8	 & 	0.000147	 & 	99,156	 & 	15	 & 	99,148	 & 	7,048,319	 & 	71.1\\
8-9	 & 	0.000132	 & 	99,141	 & 	13	 & 	99,134	 & 	6,949,171	 & 	70.1\\
9-10	 & 	0.000117	 & 	99,128	 & 	12	 & 	99,122	 & 	6,850,036	 & 	69.1\\
10-11	 & 	0.000109	 & 	99,116	 & 	11	 & 	99,111	 & 	6,750,914	 & 	68.1\\
11-12	 & 	0.000118	 & 	99,105	 & 	12	 & 	99,100	 & 	6,651,803	 & 	67.1\\
12-13	 & 	0.000157	 & 	99,094	 & 	16	 & 	99,086	 & 	6,552,704	 & 	66.1\\
13-14	 & 	0.000233	 & 	99,078	 & 	23	 & 	99,067	 & 	6,453,618	 & 	65.1\\
14-15	 & 	0.000339	 & 	99,055	 & 	34	 & 	99,038	 & 	6,354,551	 & 	64.2\\
15-16	 & 	0.000460	 & 	99,022	 & 	46	 & 	98,999	 & 	6,255,513	 & 	63.2\\
16-17	 & 	0.000577	 & 	98,976	 & 	57	 & 	98,947	 & 	6,156,514	 & 	62.2\\
17-18	 & 	0.000684	 & 	98,919	 & 	68	 & 	98,885	 & 	6,057,566	 & 	61.2\\
18-19	 & 	0.000769	 & 	98,851	 & 	76	 & 	98,813	 & 	5,958,681	 & 	60.3\\
19-20	 & 	0.000832	 & 	98,775	 & 	82	 & 	98,734	 & 	5,859,868	 & 	59.3\\
20-21	 & 	0.000894	 & 	98,693	 & 	88	 & 	98,649	 & 	5,761,134	 & 	58.4\\
21-22	 & 	0.000954	 & 	98,605	 & 	94	 & 	98,558	 & 	5,662,485	 & 	57.4\\
22-23	 & 	0.000990	 & 	98,511	 & 	98	 & 	98,462	 & 	5,563,928	 & 	56.5\\
23-24	 & 	0.000997	 & 	98,413	 & 	98	 & 	98,364	 & 	5,465,466	 & 	55.5\\
24-25	 & 	0.000982	 & 	98,315	 & 	97	 & 	98,267	 & 	5,367,101	 & 	54.6\\
25-26	 & 	0.000960	 & 	98,219	 & 	94	 & 	98,171	 & 	5,268,835	 & 	53.6\\
26-27	 & 	0.000942	 & 	98,124	 & 	92	 & 	98,078	 & 	5,170,663	 & 	52.7\\
27-28	 & 	0.000936	 & 	98,032	 & 	92	 & 	97,986	 & 	5,072,585	 & 	51.7\\
28-29	 & 	0.000947	 & 	97,940	 & 	93	 & 	97,894	 & 	4,974,599	 & 	50.8\\
29-30	 & 	0.000974	 & 	97,847	 & 	95	 & 	97,800	 & 	4,876,705	 & 	49.8\\
30-31	 & 	0.001008	 & 	97,752	 & 	98	 & 	97,703	 & 	4,778,906	 & 	48.9\\ \hline
    \end{tabular}
\end{center}
\end{table}
\begin{table}{}
\begin{center}
    \begin{tabular}{|l|c|c|c|c|c|c|}
    \hline
    Age & Probability of & Number & Number  & Person-years & Total number & Expec.   \\
    & dying between & surviving & dying  ages  & lived between  & of person-years & of life \\
    & ages $$ to $x+1$ & to age $x$ & between ages & ages $x$ to $x+1$ & lived above & at age $x$ \\
    &               &          &  $x$ to $x+1$        &               & age $x$           &            \\
    \hline
31-32	 & 	0.001046	 & 	97,654	 & 	102	 & 	97,603	 & 	4,681,203	 & 	47.9\\
32-33	 & 	0.001097	 & 	97,551	 & 	107	 & 	97,498	 & 	4,583,600	 & 	47.0\\
33-34	 & 	0.001162	 & 	97,444	 & 	113	 & 	97,388	 & 	4,486,102	 & 	46.0\\
34-35	 & 	0.001244	 & 	97,331	 & 	121	 & 	97,271	 & 	4,388,715	 & 	45.1\\
35-36	 & 	0.001336	 & 	97,210	 & 	130	 & 	97,145	 & 	4,291,444	 & 	44.1\\
36-37	 & 	0.001441	 & 	97,080	 & 	140	 & 	97,010	 & 	4,194,299	 & 	43.2\\
37-38	 & 	0.001567	 & 	96,940	 & 	152	 & 	96,864	 & 	4,097,289	 & 	42.3\\
38-39	 & 	0.001714	 & 	96,788	 & 	166	 & 	96,705	 & 	4,000,424	 & 	41.3\\
39-40	 & 	0.001874	 & 	96,623	 & 	181	 & 	96,532	 & 	3,903,719	 & 	40.4\\
40-41	 & 	0.002038	 & 	96,442	 & 	197	 & 	96,343	 & 	3,807,187	 & 	39.5\\
41-42	 & 	0.002207	 & 	96,245	 & 	212	 & 	96,139	 & 	3,710,844	 & 	38.6\\
42-43	 & 	0.002389	 & 	96,033	 & 	229	 & 	95,918	 & 	3,614,705	 & 	37.6\\
43-44	 & 	0.002593	 & 	95,803	 & 	248	 & 	95,679	 & 	3,518,787	 & 	36.7\\
44-45	 & 	0.002819	 & 	95,555	 & 	269	 & 	95,420	 & 	3,423,108	 & 	35.8\\
45-46	 & 	0.003064	 & 	95,285	 & 	292	 & 	95,139	 & 	3,327,688	 & 	34.9\\
46-47	 & 	0.003322	 & 	94,993	 & 	316	 & 	94,836	 & 	3,232,548	 & 	34.0\\
47-48	 & 	0.003589	 & 	94,678	 & 	340	 & 	94,508	 & 	3,137,713	 & 	33.1\\
48-49	 & 	0.003863	 & 	94,338	 & 	364	 & 	94,156	 & 	3,043,205	 & 	32.3\\
49-50	 & 	0.004148	 & 	93,974	 & 	390	 & 	93,779	 & 	2,949,049	 & 	31.4\\
50-51	 & 	0.004458	 & 	93,584	 & 	417	 & 	93,375	 & 	2,855,270	 & 	30.5\\
51-52	 & 	0.004800	 & 	93,167	 & 	447	 & 	92,943	 & 	2,761,895	 & 	29.6\\
52-53	 & 	0.005165	 & 	92,719	 & 	479	 & 	92,480	 & 	2,668,952	 & 	28.8\\
53-54	 & 	0.005554	 & 	92,241	 & 	512	 & 	91,984	 & 	2,576,472	 & 	27.9\\
54-55	 & 	0.005971	 & 	91,728	 & 	548	 & 	91,454	 & 	2,484,487	 & 	27.1\\
55-56	 & 	0.006423	 & 	91,181	 & 	586	 & 	90,888	 & 	2,393,033	 & 	26.2\\
56-57	 & 	0.006925	 & 	90,595	 & 	627	 & 	90,281	 & 	2,302,145	 & 	25.4\\
57-58	 & 	0.007496	 & 	89,968	 & 	674	 & 	89,630	 & 	2,211,864	 & 	24.6\\
58-59	 & 	0.008160	 & 	89,293	 & 	729	 & 	88,929	 & 	2,122,234	 & 	23.8\\
59-60	 & 	0.008927	 & 	88,565	 & 	791	 & 	88,169	 & 	2,033,305	 & 	23.0\\
60-61	 & 	0.009827	 & 	87,774	 & 	863	 & 	87,343	 & 	1,945,136	 & 	22.2\\
61-62	 & 	0.010831	 & 	86,911	 & 	941	 & 	86,441	 & 	1,857,793	 & 	21.4\\
62-63	 & 	0.011872	 & 	85,970	 & 	1021	 & 	85,460	 & 	1,771,352	 & 	20.6\\
63-64	 & 	0.012891	 & 	84,949	 & 	1095	 & 	84,402	 & 	1,685,892	 & 	19.8\\
64-65	 & 	0.013908	 & 	83,854	 & 	1166	 & 	83,271	 & 	1,601,490	 & 	19.1\\
65-66	 & 	0.015003	 & 	82,688	 & 	1241	 & 	82,068	 & 	1,518,219	 & 	18.4\\
66-67	 & 	0.016267	 & 	81,448	 & 	1325	 & 	80,785	 & 	1,436,151	 & 	17.6\\
67-68	 & 	0.017699	 & 	80,123	 & 	1418	 & 	79,414	 & 	1,355,366	 & 	16.9\\ \hline
\end{tabular}
\end{center}
\end{table}

We compute the probabilities that the pension survives the pensioner, for the retirements ages 67 and 72, for different pension/consumption ratios.
\renewcommand{\arraystretch}{1.1}
\begin{table}
\begin{center}
  \begin{tabular}{|c|c| }
    \hline
    Initial Pension                        & Chance to Die \\
    $\overline{\mbox{Yearly Consumption}}$ & Before Pension is Consumed \\ \hline\hline
    7.5 Years            &19.18\%\\ \hline
    10 Years             &28.65\%\\ \hline
    12 Years             &54.70\%\\ \hline
    12.5 Years           &60.29\%\\ \hline
    15 Years             &67.43\%\\ \hline
    16.25 Years          &72.62\%\\ \hline
  \end{tabular}
\end{center}
  \caption{\small Pensions surviving 67 years old pensioners}
\end{table}
\begin{table}
\begin{center}
  \begin{tabular}{|c|c|}
    \hline
    Initial Pension                        & Chance to Die \\
    $\overline{\mbox{Yearly Consumption}}$ & Before Pension is Consumed \\ \hline\hline
    7.5 Years                    &28.18\%\\ \hline
    10 Years                    &40.93\%\\ \hline
    12 Years                    &60.70\%\\ \hline
    12.5 Years                    &65.39\%\\ \hline
    15 Years                    &78.13\%\\ \hline
    16.25 Years                    &87.78\%\\ \hline
  \end{tabular}
\caption{\small Pensions surviving 72 years old pensioners}
\end{center}
\end{table}

\newpage
\section{Summary and conclusions}
%performance
We developed here a stochastic model of the random environment of a pension plan that is invested in the stock market. The model is based on historical stock market data. Data analysis confirms the EMH which implies that there is no benefit in the pension fund setting up a super fund that owns "good" companies lock stock and barrel. We find that CPI-adjusted salaries decrease over time, while CPI-adjusted market returns drift upwards. Assuming that past market dynamics persists in the future, our model estimates that the pensioner is likely to accumulate $7.5$ times his initial annual salary over 40 years of pension savings. Assuming 10\% salary contribution, this means that the pension portfolio CPI-adjusted average annual return is 2.85\%. We find that there is only $19.18\%$ chance for this pension to be sufficient, assuming retirement age at 67. In other words, a salaried employee, who worked his entire life without receiving any substantial promotions, bonuses or extra incomes, has about 80\% chance to live his last 10-15 years in poverty. Raising the retirement age to 72 is expected to bring these chances down to about 70\%. These results shed further light on the imminence and significance of the pension poverty problem.

%economic implications
It is not clear from the model and its analysis what is an investment strategy that the tax payer can adopt to insure pensions.  Clearly, owning companies on a national scale might form a centrally planned economy, in which the government owns a portion of the means of production. This could lead to economic inefficiencies observed in socialized economies in the past. Investing pension funds in the economy can be bolstered by the tax payer by expanding it through investments in infrastructure such as large scale public utilities, mainly in education and continuing education. Tax money is funnelled into the private sector in the USA by government contracts for projects and services, but not necessarily into direct subsidies for investors or in the form of preferential tax breaks. The latter is a common practice in Israel, though.

To conclude, there is an imminent urge for the structuring of a long-term investment scheme that secures the pensioner welfare, yet it is a complicated, large-scale problem. Based on numerical results, a symptomatic treatment of the pension problem can be achieved by raising the retirement age and increasing contributions. However, in order to achieve a systematic solution, researchers from the entire scientific spectrum need to contribute to the effort.

\end{document}